# Electrooptics of mm-scale polar domains in the ferroelectric splay nematic phase

Nerea Sebastián,[1*] Richard J. Mandle,[2] Andrej Petelin[1,3], Alexey Eremin[4] and Alenka Mertelj[1*]

[1]Jožef Stefan Institute, P.O.B 3000, SI-1000 Ljubljana, Slovenia

[2] School of Physics and Astronomy, University of Leeds, Leeds, UK, LS2 9JT

[3]Faculty of Mathematics and Physics, University of Ljubljana, Slovenia

[4]Department of Nonlinear Phenomena, Institute for Experimental Physics Otto von Guericke University Magdeburg Universitätsplatz 2, 39106 Magdeburg (Germany)

Corresponding author:

Nerea Sebastián: nerea.sebastian@ijs.si

Alenka Mertelj: alenka.mertelj@ijs.si

**Abstract**

The recent discovery of the ferroelectric splay nematic phase has opened the door to experimental investigation of one of the most searched liquid crystal phases in decades, with high expectations for future applications. However, at this moment, there are more questions than answers. In this work, we examine the formation and structure of large polar nematic domains of the ferroelectric splay nematic material RM734 in planar liquid crystals cells with different aligning agents and specifications. We observe that confining surfaces have a strong influence over the formation of different types of domains, resulting in various twisted structures of the nematic director. For those cells predominantly showing mm-scale domains, we investigate the optical and second harmonic generation switching behaviour under applications of electric fields with a special focus on in-plane fields perpendicular to the confinement media rubbing direction. In order to characterize the underlying structure, the polar optical switching behaviour is reproduced using a simplified model together with Berreman calculations.

**Introduction**

A ferroelectric nematic phase is of great fundamental and practical interest. The simplest of the nematic (*N*) phases, that formed by achiral elongated molecules, does not show any positional order, but on average, molecules orient along the same direction. Such direction, called the director, is represented by a unit vector **n**, which in the absence of polarity shows head-to-tail symmetry, $\mathbf{n(r)} = -\mathbf{n(r)}$. The possibility of a ferroelectric nematic phase, where the head-to-tail symmetry is broken, was already envisioned by Born.[1] Over the years, was followed by several theoretical works contemplating the possibility of polar nematic order in discotic liquid crystals (LC),[2,3] or the occurrence of a splayed-polar nematic phase for pear-shaped molecules.[4–6]

In the past few years, we have reported a modulated ferroelectric splay nematic phase ($N_s$), occurring in a material made of slightly wedge-shaped molecules carrying a large dipole moment of ~11.4 D (*RM734*, see Fig. 1).[7–9] On cooling, RM734 exhibits two distinct nematic phases, separated by a weakly first-order transition.[10,11] In the high temperature *N* phase, the splay elastic constant rapidly decreases when approaching the *N*-$N_s$ transition, resulting in a significant pre-transitional behaviour, manifested by strong splay orientational fluctuations. Instability towards splay orientational deformation, arising from the flexoelectric coupling, drives the transition to a periodically splayed structure. The divergent behaviour of the electric susceptibility shows that it is a paraelectric-ferroelectric phase transition, which, due to the flexoelectric coupling, is accompanied by an orientational ferroelastic transition. The polarity of the splay phase was further proven by second harmonic generation (SHG) imaging, showing that the splay modulation period at the transition is of the order of 5-10 µm. Modulation in the µm range was also reported in a chemically induced $N_s$ phase, in the mixture of two materials which neither of them exhibits the $N_s$ phase: a non-mesogenic bent-core material, which retains some of the features of RM734, and a nematic phase made of molecules analogous in structure to *RM734* but with longer terminal chains.[12] Such mixture exhibits the $N_s$ phase at room temperature (8 degrees below the *N*-$N_s$ transition), for which periodicities ~9 µm have been observed optically. One should note that, particularly in confined geometries and deeper in the ferroelectric splay nematic phase, the periodic structure is not the only possibility. Structures can be more complicated and strongly influenced by the boundary conditions, as will be explored in this contribution. In a very recent work the spontaneous polarization of RM734 has been estimated to be $P \sim 6\ \mu C/cm^2$ by means of field-induced current measurements in thin cells.[13] Consistent with the observations in both systems are those results for the spontaneous polarization ($P \sim 4.4\ \mu C/cm^2$) and SHG in the MP2 phase initially reported by Nishikawa et al.[14] for a slightly wedge-shaped 1,3-dioxane-based molecule carrying also a large dipole moment (~9.4 D). Very recently Li et al.[15] synthesized a large number of elongated molecules with dipole moments ranging from 4 to 13D, some exhibiting stable $N_s$ phase, some metastable $N_s$ and many others no $N_s$ phase at all. Utilizing machine-learning analyses they show how molecular parameters such as molecular dipole moment, molecular aspect ratio, molecular length or dipole angle with the molecular axes influence the stabilization of the polar nematic phase. Interestingly, additionally to a large dipole moment, it appears that a moderate angle (20°) between the molecular axis and the dipole moment is required.

The *N*-$N_s$ phase transition has been described by a Landau-de Gennes type of phenomenological theory,[7,8,16] contemplating the flexoelectric coupling between splay deformation and the polar order, similarly to the model for the transition between the nematic and the twist-bend nematic phase of Shamid et al.[17] Very recently, a 2D splay modulation of the director field, together with a 2D modulation of the polar order, has been predicted by Rosseto and Selinger,[18] to be the more likely structure for the $N_s$ phase. In their calculations, they show that the 2D structure is in a general energetically more favourable than the 1D modulation, which would only be more favourable in a narrow interval between the uniform nematic and the 2D splay phase. Although this model is good for the description of the phase transition, deeper in the splay phase, the simple description is no longer sufficient, as will be shown below. Because it is not possible to fill the space with uniform splay, a modulated splay structure resembling alternating ferroelectric domains could be a ground state at a given temperature interval. Splay deformation in such a phase would be visible by polarizing optical microscopy (POM), if the splay deformation lays in the plane of the layer, a situation, which is typically disfavoured by the homogeneous surface anchoring conditions. The splay deformation would also cause a small optical biaxiallity, but because of its smallness compared to birefringence, it would be very difficult to observe it optically. However, the tendency to splay would manifest in the elastic properties, in particular, in the softening of the splay elastic constant, which is exactly what is observed at the phase transition to the ferroelectric phase.[7,8] The large modulation period observed below the phase transition to the ferroelectric phase, comparable to the standard thicknesses of LC cells (5-20 μm), indicates that the splay deformation is small, and that it is reasonable to anticipate that in a confined system, e.g. such as a typical LC cell, more frustrated structures with similar energies will exist. These structures will be strongly affected by the boundary conditions, and which of these structures will be observed will depend on the history of the sample, e.g. exposure to the external fields, cooling rates etc.

In this work we aim to contribute to forming a base of experimental observations for the ferroelectric splay nematic phase which will help to further advance in the understanding of this novel nematic phase and to inform the development of theoretical models. For this purpose, we first analyse the formation of different domain structures immediately following the $N$-$N_S$ phase transition. Furthermore, we examine their electro-optic behaviour under the application of electric fields with special emphasis on in-plane electric fields perpendicularly to **n**, both using polarizing optical microscopy (POM) and by SHG imaging (see Methods). We investigate the switching behaviour in the predominant type of domains observed and reproduce it by means of a simplified model and Berreman calculations. The proposed twist structure is finally faced with results from cross-Dynamic Differential Microscopy (c-DDM).[19] This paper is thus organized as follows. Material and results are thoroughly presented in the Results section. Careful examination of them is presented in the Discussion section. Additional material, as the videos corresponding to the results discussed here, is provided in the Electronic supplementary information (ESI).

## Results

### Material

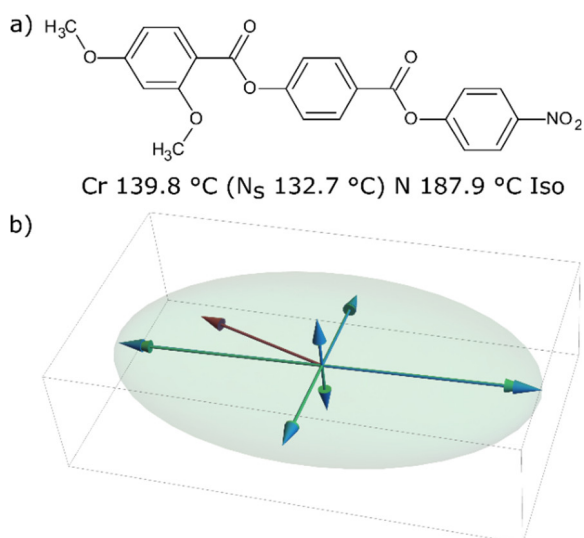

Fig. 1. (Color online) a) Molecular structure of RM734 and its phase transitions. b) Representation of dipole moment (red arrow), polarizability tensor at 800nm (green axis and ellipsoid) and axis of the static polarizability tensor (blue arrows).

The structure of the liquid-crystalline material RM734 (4-((4-nitrophenoxy)carbonyl)phenyl-2,4-dimethoxybenzoate) and the phase sequence are presented in Fig. 1.a. On heating, the crystalline phase melts directly into the nematic phase at 139.8 °C and transforms into the isotropic liquid phase at 187.9 °C. On cooling, *RM734* exhibits the isotropic (*I*) to nematic (*N*) phase transition followed by nematic to ferroelectric splay nematic ($N_S$) transition at 132.7 °C, which crystallizes around 90 °C, temperature which can vary depending on the cooling rate. The synthesis of *RM734* is described in ref. 10. *RM734* has a very large dipole moment (~11.4 D) as calculated at the M06HF-D3/aug-cc-pVTZ level of DFT, at an angle of 18.3° with respect to the molecular axis and is therefore mainly longitudinal.[20] Fig.1.b shows the molecular dipole moment (red arrow) and the polarizability tensor (green arrows and ellipsoid) as calculated at the M06HF-D3/aug-cc-pVTZ level of DFT.[20] Whereas the molecular dipole moment forms an angle with respect to the molecular axis, the long axis of polarizability tensor is mostly defined by it.

The formation of domains in the $N_S$ phase was explored in commercial EHC cells of different thicknesses (KSRP D-type of 5, 9, and 25 µm) and with planar aligning layer (aligning agent polymide LX-1400 from Hitachi-Kasei and antiparallel rubbing). Response to applied electric fields was studied in commercial in-plane switching (IPS) liquid crystal cells purchased from Instec, with an electrode width and gap between electrodes of 15 µm and a cell thickness of 9 µm, and with planar alignment parallel to the electrodes (aligning agent KPI-300B, antiparallel rubbing). The IPS operation mode is illustrated in Fig. 4.b. All such measurements were performed after cooling from the $N$ phase. Additionally, $N$-$N_S$ transition was also inspected in Instec IPS cells, equivalent to the former, but with the planar rubbing perpendicular to the electrodes (same aligning agent and antiparallel rubbing).

**$N$-$N_S$ transition and domains**

Regardless of the cell thickness, on cooling from the $N$ phase (Fig. 2.a), the $N$-$N_S$ transition is characterized by the observation, just above the transition, of initial freezing of the characteristic nematic flickering (director fluctuations) followed by strong fluctuations destabilizing the homogeneous orientation of **n**, characterized by a striped texture (Fig. 2.b). As shown in reference 8 and here below, such stripes show a strong alternating SHG signal. On further cooling, a homogeneous texture is restored together with the flickering (Fig. 2.b) and afterwards a bright line front, usually starting on defect points, travels across the sample (Fig. 2.c). When two of these fronts come together, they either annihilate or, as in the case shown in ESI.Video1, result in a domain wall (Fig. 2.d). Such walls, which are "soft", usually reposition over time and deform under the application of electric fields, as will be shown below. Interestingly, when the wall travels parallel to the rubbing direction, it exhibits a strong tendency to show a sierra-shape configuration, with the walls at angles between 50 and 70 degrees from the rubbing direction (Fig. 2). When directed perpendicular to the rubbing direction, they slightly bend but run over long distances without drastic deformations. When observed with the cell's rubbing direction along the polarizers, the $N_S$ phase before the wall propagation shows very good extinction (Fig. 2.e). However, after wall propagation dark extinction is lost for thin cells. Here, it should be mentioned that the final state and homogeneity of the sample depends strongly on the used LC cell. In EHC planar cells clean propagation of the wall leads to very large domains spanning up to several millimetres, sometimes even covering almost the full cell. Same behaviour is found for cells with smaller (5.4 µm) or larger (22 µm) thickness (Fig. ESI.1)

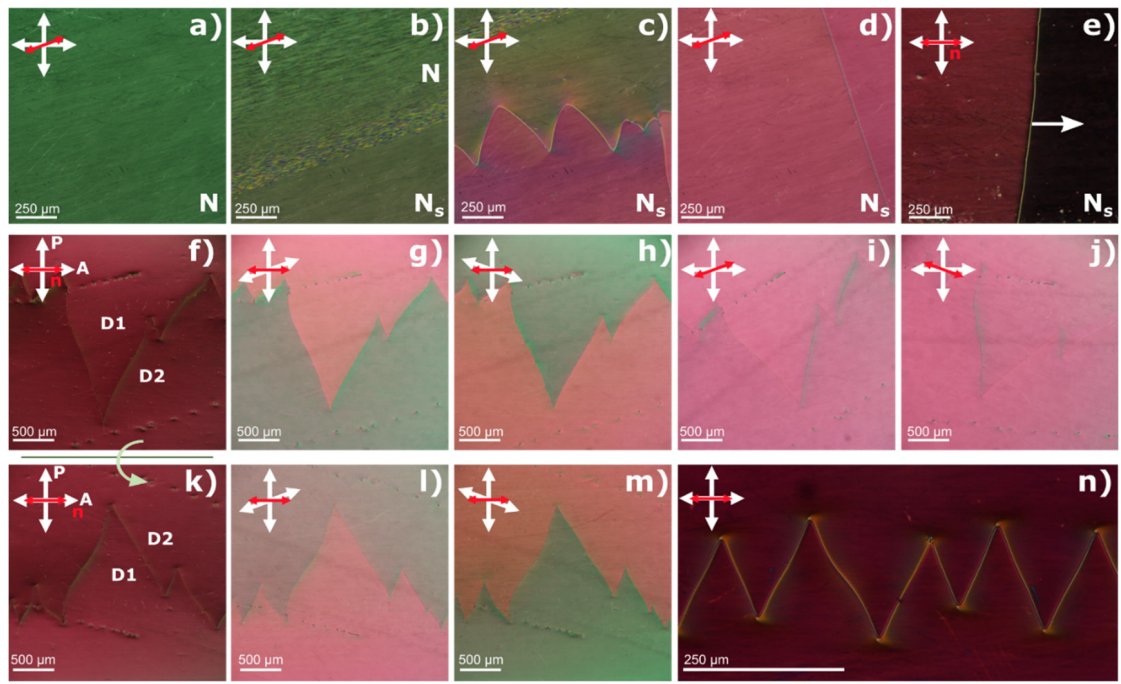

Fig. 2. (Color Online) $N$-$N_S$ transition and mm-size domains (a-e) Snapshots across the $N$-$N_S$ transition. From left to right: (a) uniform $N$ phase, (b) destabilization of the homogeneous n orientation characterized by a stripped texture, (c) propagation of a π wall and (d) final $N_S$ texture. (e) π -wall propagation for the sample in the extinction position. (f) Texture of the $N_S$ phase of RM734 under crossed polarizers in extinction position and (g-h) by uncrossing the analyser by 20° clockwise and anticlockwise, indicating domains of opposite handedness separated by "soft" walls. (i-j)Rotation of the sample between crossed polarizer's reveals symmetric behaviour for rotations of 20 degrees clock and anticlockwise. (k-m) Flipping the sample upside-down around the cell rubbing direction evidences the same optical activity for domains D1 and D2 under uncrossing the analyser. n) Typical sierra-shape configuration of the "soft" walls separating domains. Cell thickness is 8.3 µm for images (a-m), 5.4 µm for (n) and red allow indicates the rubbing direction.

In either standard or IPS Instec cells, which use a different aligning agent to the EHC cells, in some areas some "lens-structures" develop before the front propagation (Fig. ESI.2 and ESI.Video 2). Such structures can be small, like reported in ref. 13, or quite large as we usually found in some areas of IPS cells (up to several hundreds of µm). When the final structural relaxation propagates through the cell, the characteristic front does not travel through these structures, but the wall divides and surrounds them giving rise to a bright domain wall which is pinned to the surface (Fig. ESI.2 for localized structures and ESI.Video 2 for larger areas). Such walls, which are "hard", stay fixed in place and do not deform under electric fields. They can be localized or cover the whole cell and can be found both, in freshly filled cells (Fig. ESI.3.f), indicating a precondition of the cell, or in cells where they were initially absent after application of electric fields. In either case, heating the sample back to the $N$ phase melts them but on cooling back to the $N_S$ phase the "hard" walls tend to appear in the same areas, indicating a strong surface memory effect.

Further information can be obtained from POM observations. Focusing attention on those domains divided by "soft" walls, it can be seen that such domains do not show extinction when aligning the cell rubbing direction with the analyser as shown in Fig. 2.e and Fig. 2.f. When uncrossing of the analyser by a small angle (20°) in opposite directions, opposite optical activity is revealed suggesting the presence of a twisted structure with opposite handedness (see Fig. 2.g and Fig. 2.h). Rotation of the sample anticlockwise and clockwise with respect to the crossed polarizers reveals no textural differences between both domains in this geometry. Up-down rotation of the sample around the cell´s rubbing directions shows that each domain shows the same optical activity on uncrossing the analyser in the same direction, indicating the presence of just one twist handedness through the domain. Same studies on those domains enclosed by "hard" walls (Fig. ESI.3), reveal that although uncrossing the analyser results in similar observations, rotation of the sample clockwise or anticlockwise shows opposite colour texture (Fig. ESI.3. d-e and Fig. ESI.3. d-e and g-h), evidencing that the twist structure inside these domains should be different than that of the predominant ones.

Finally, we checked the behaviour of the $N$-$N_S$ transition in IPS cells equivalent to the previous ones, but, in which the aligning rubbing direction (same aligning agent and antiparallel rubbing, see Materials) is directed perpendicularly to the electrodes. Interestingly, the $N$-$N_S$ transition manifests differently in this case. Destabilization of the homogeneous nematic director and the striped texture are followed in this case by the formation of large elongated structures, covering the whole area of the cell, which flow, join annihilate and slowly stabilize into smaller domains embedded in a larger domain (Fig. ESI. 4 and ESI. Video 3). Structural relaxation as observed in the previous cells (Fig. 2.d and e) does not occur. In the present case, sudden changes in the walls separating the domains can be observed, propagating from one end to the other (ESI. Video 3). The final texture appears uniform with embedded domains. Aligning the cell rubbing direction with the crossed polarizers reveals a much better extinction than in the previous cells (Fig. ESI.5.a). When, uncrossing polarizers in extinction position and rotating the sample between crossed polarizers symmetric optical behaviour can be observed, changing inside the smaller domains, but also varying in the matrix domain at long distances (Fig. ESI. 5.(b-e)). This indicates, that the director structure might not be as uniform as would be assumed at first sight.

In this contribution, due to their prevalence in the different observed cells, we mainly focus on the electro-optic behaviour of the complementary domains divided by the soft walls described in this section (Fig. 2).

**Application of out of plane fields**

As mentioned previously, when cooling from the N phase into the $N_S$, very large domains of several millimetres are obtained (Fig. 2 and Fig. 3.a-c). Application of out of plane electric fields shows a variety of behaviours, whose in-depth analysis is not the scope of this paper. There are however several features that are worth mentioning. In EHC cells (5.4 µm), under application of low frequency (1 Hz) square-wave voltages and at low amplitudes (up to Vp=1 V) the optical texture remains unchanged, walls slightly deform and domains blink in a non-uniform manner, strongly depending on surface defects. For higher voltages (Vp=1.5 V, see Fig. 3.d), the domain walls start to strongly deform, with those running parallel to the rubbing direction adopting the sierra-shape configuration mentioned before (see area surrounded by the ellipsoid in Fig. 3.d), while those running perpendicularly to the rubbing direction retaining their shape. In some spots in the middle of domains where prior to field application no defect or dust was present, line defects initially grow which gradually

"open" giving rise to new domains. Fig. 3.e and Fig. 3.f shows the same sample area as in Fig. 3.a, after the field was switched off, evidencing that the initially large domains have broken down into smaller domains by the emergence of new ones, but with similar optical activity as the originating ones. New domain walls formed in this way show a stronger tendency for the sierra-shape configuration, with much shorter periodicity. The application now of square-wave of low amplitudes and different offsets reveals a second interesting finding observed repeatedly in these domains. When voltage is alternated between –Vp and Vp domains remain "unchanged" just blinking as described before (Fig. ESI.6.a and ESI.Video.4). If voltage switches from 0 to 2Vp one of the domains (D1) starts to progressively shrink (or equivalently domain D2 starts to grow), showing that such field polarity is not favourable for this domain (Fig. ESI.6.b and ESI.Video.4). If the field is switch back to alternate between –Vp and Vp domain walls stop and stay in place. Finally, if now the voltage switches from –2Vp to 0, those domains previously shrinking, start to gradually grow again (Fig. ESI.6.c and ESI.Video.4) indicating that in this case, the field polarity is favourable within the domain.

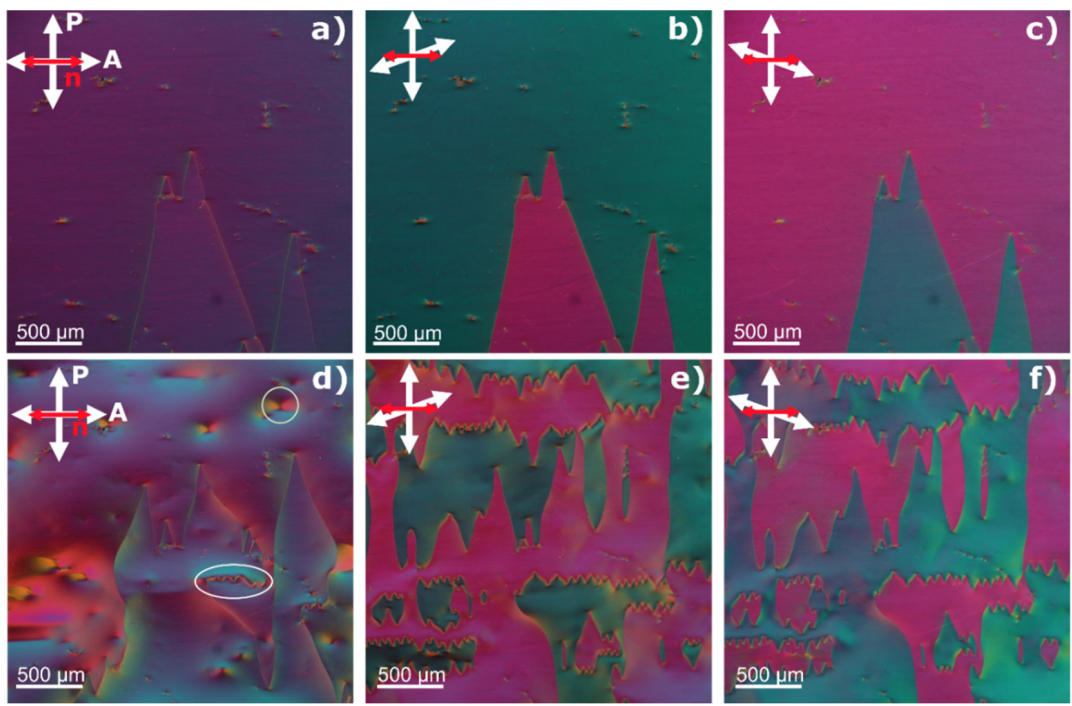

Fig. 3. (Color online) Domain division/creation by application of out of plane fields, in an EHC 5.4 µm cell at T=120 °C. Before application of field between crossed polarizers (a), and with analyser uncrossed (b-c). d) Texture observed during the application of 1Hz square-wave field of Vp= 1.5 V. (e-f) Final state of the sample after the field is switched off.

**In-plane electro-optic behaviour**

**Pulses**

We investigated the electro-optic response to in-plane electric fields of the $N_s$ phase at a temperature well below the transition (120 °C, $T_{N-Ns}$-T=12), by means of polarizing optical microscopy (POM) as described in the Methods section at the end. To minimize the effect of mobile charges, we applied a voltage pulse sequence (see Fig. 4.c) consisting of two opposite 50 ms pulses separated by 450 ms and with a repetition rate of 1 second, with peak voltages between Vp=30 mV and Vp=1.2 V ($E$ fields between 2 V/mm and 80 V/mm in the plane between the electrodes). Experiments were performed for three different angles ($\phi$ = 0°, 20° and 45°) between the rubbing directions and the crossed polarizers and the sequence corresponding to one second period was then reconstructed prior analysis (ESI.Video.5). For the rubbing direction along the crossed polarizers, the typical electro-optic response is given in Fig. 4.a, corresponding to snapshots for three different values, but at the same polarity, of the applied electric field for two domains separated by a "soft wall". Similarly, in Fig. 4.d, the optical response at different times of the pulse sequence is shown for a given field. There are two significant characteristics to be noted. Firstly, the optical response to the test fields inside a given domain is non-symmetric with respect to the field direction. Secondly, the optical response is equivalent, but opposite with respect to the field polarity at both sides of the "soft wall" dividing the domains. This is clearly shown in the mean intensity profile plot in Fig. 4.e and the time dependence mean intensity plots shown in Fig. 5.a and Fig.ESI.7 for different areas in the texture corresponding to different field directions, in-plane (between electrodes) and out of plane (above electrodes). Inside a given domain, for voltages below Vp= 600 mV, the transmitted intensity between crossed polarizers increases for a given field direction during the pulse duration, while decreases when the field is in the opposite direction as shown in Fig. 5.a. Given the alternating field distribution of the IPS cells as sketched in Fig. 4.b, the described behaviour results in alternating switching inside a domain. The next domain, on the other hand, shows the same response but reversed with respect to the voltage peak sign (Fig. 4.(d,e), Fig. 5.a and Fig.ESI.7). When varying the field amplitude, for a given gap within electrodes, transmitted intensity increases steadily for one field direction, while decreases, reaches a minimum and starts increasing, for the opposite field direction (Fig. 5.b, Fig. 6.b and Fig.ESI.7).

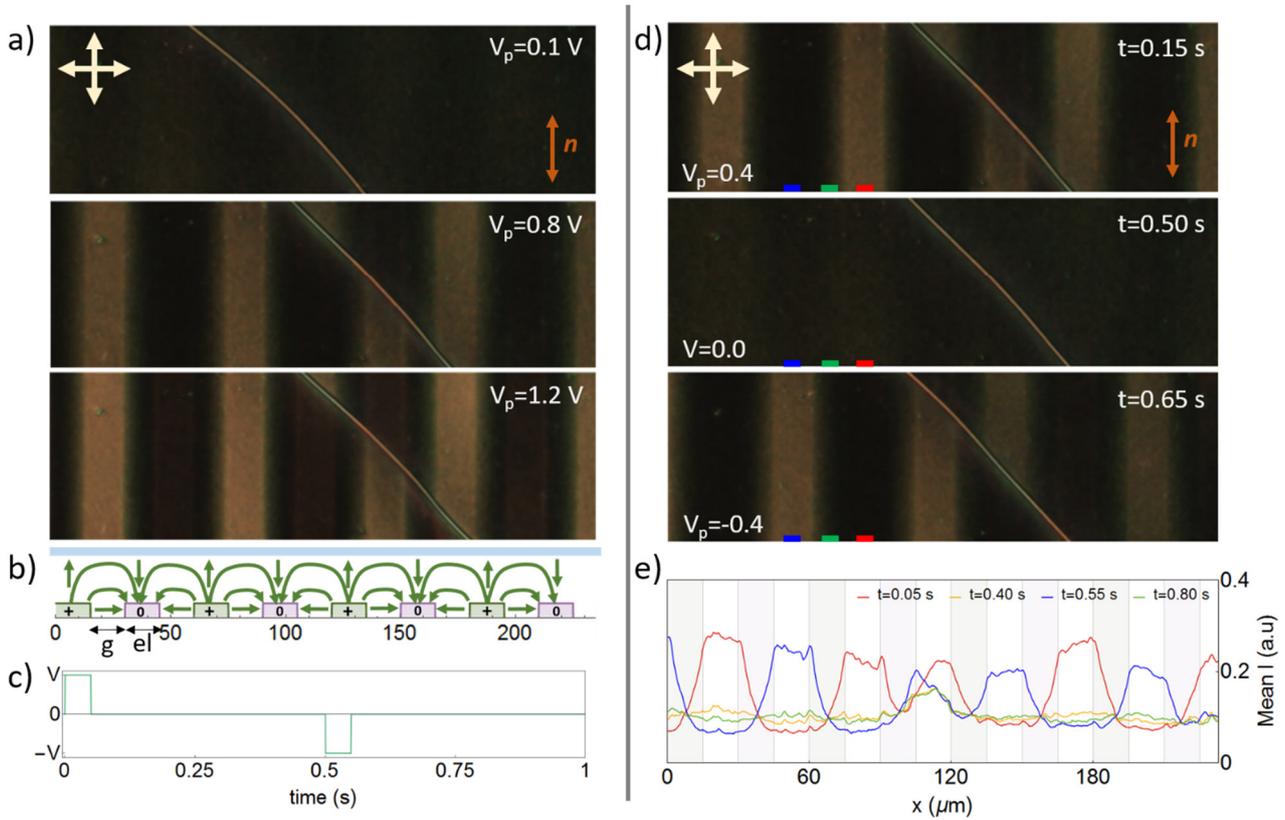

Fig. 4. (Color online) a) Polarizing optical microscopy (POM) images showing in-plane switching of the $N_s$ phase at 120 °C ($T_{N-Ns}$-T=12) for two opposite domains separated by a domain wall at different $V_p$. b) Schematic of the corresponding section of the cell geometry, where green arrow represents the electric field vector for a given polarity of the field. Both, the width of the electrode (e) and of the gap (g), corresponds to 15 μm. c) Applied pulse sequence, with two opposite pulses of 50 ms, separated by 0.5 s. d) POM snapshots of the switching sequence at three different times corresponding to the pulses peaks and 0 field. e) Intensity profile over the direction perpendicular to the electrodes for the snapshots in (d). Time has been shifted 0.1 seconds for correspondence with Fig. 5 and better visualization.

In the case of the electro-optic response recorded for the rubbing direction at 20° with respect to the crossed polarizers, results are comparable (Fig. ESI.8). Transmitted intensity at zero field ($I_0$) increases as expected from $I = I_n \sin^2 \phi \sin^2 (\pi \Delta n\, d/\lambda)$. Normalized intensities over reorientation are qualitatively similar to what is observed for $\phi = 0°$, while the larger $I_0$ value results in lower contrast and the minimum on the normalized mean intensity slightly shifts towards lower fields. However, when we observe the electro-optic response at $\phi = 45°$ (Fig. ESI.9), the difference between both field directions in the same electrode gap becomes nearly indistinguishable. Additionally, the minimum of the transmitted intensity shifts nearly to zero field, but of the opposite sign from the scenario for $\phi = 0°$ and $\phi = 20°$ (Fig. 6). Finally, less noticeable changes in the transmitted intensity can be detected on top of the electrodes, however the non-uniform field profile together with the penetration of the signal coming from the large reorientation in-between the electrodes restricts the analysis.

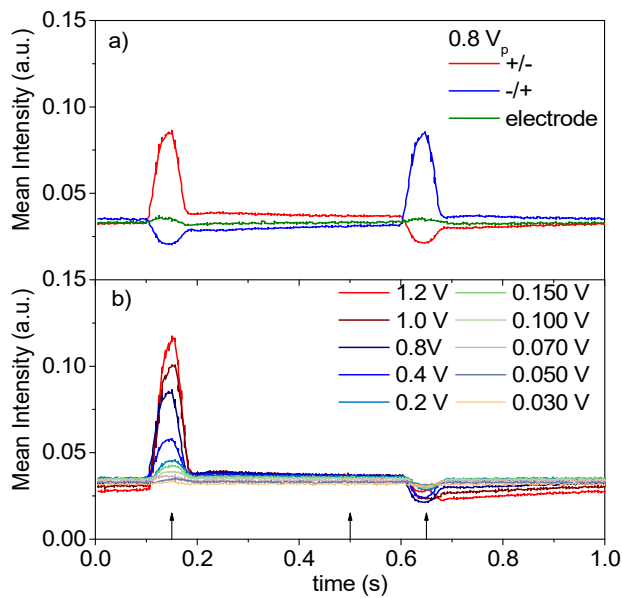

Fig. 5. (Color online) Time dependence over the switching cycle of the transmitted intensity for the cell oriented with the rubbing direction along the analyser in the $N_s$ phase at 120 °C ($T_{N-Ns}$=12). a) For a given $V_p$ for the three areas marked in Fig. 2 within the same domain: red (one field direction), blue (opposite field direction) and green (over the electrode). b) For the area marked as red at different field strengths. Time has been shifted 0.1 seconds for better visualization.

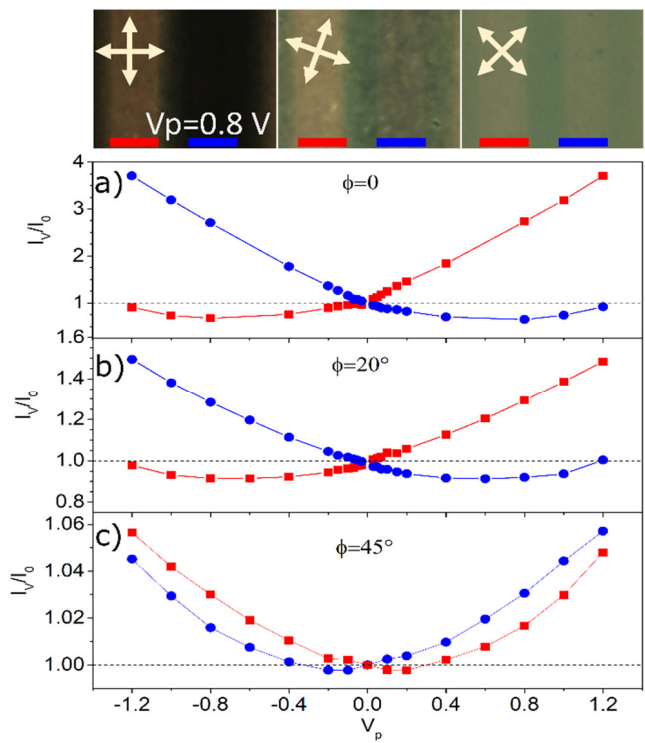

Fig. 6. (Color online) Normalized mean intensity ($I_V / I_0$) over reorientation for two areas within a domain corresponding to opposite fields, marked by the red and blue bar in the top POM images. For rubbing direction at a) 0°, b) 20° and c) 45° with respect to the crossed polarizers.

**Square-wave**

For comparison with the electro-SHG experiments that are detailed below, the electro-optic response was also studied under the application of square-wave 1 Hz driving fields, for $\phi = 0°$ (Fig. 7 and Fig. ESI.10). Results are qualitatively similar to those obtained with the pulse sequence, where, for a given domain, one field direction results in the increase of transmitted intensity while the opposite field direction leads to a decrease, clearly observed in Fig. 7.b. Due to the time dependent voltage profile, where $-V_p$ and $V_p$ are applied for sufficiently long times, mobile charges, e.g. ionic impurities in the LC, have time to redistribute, which results in screening effects on the applied fields and of the polarization charges. Such effects, lead to longer equilibration times for the electro-optic response. Comparison of the normalized mean intensity $I_V / I_0$ at different times with that obtained from the application of short pulses (Fig. ESI.11) show the reason for the usage of the latter. In experiments for which short pulses were employed, reorientation is almost, although not fully, completed and in this case the method is better for probing the equilibrium structure and in the case of square-wave, ion dynamics would need to be taken into account.

Application of square-wave also revealed a limiting voltage above which no equilibrium structure is achieved. At Vp=1.4 V structure starts to destabilize, as seen in Fig. ESI.12, eventually leading to the onsite creation of new walls and domains. Such studies although interesting, are out of the scope of this paper.

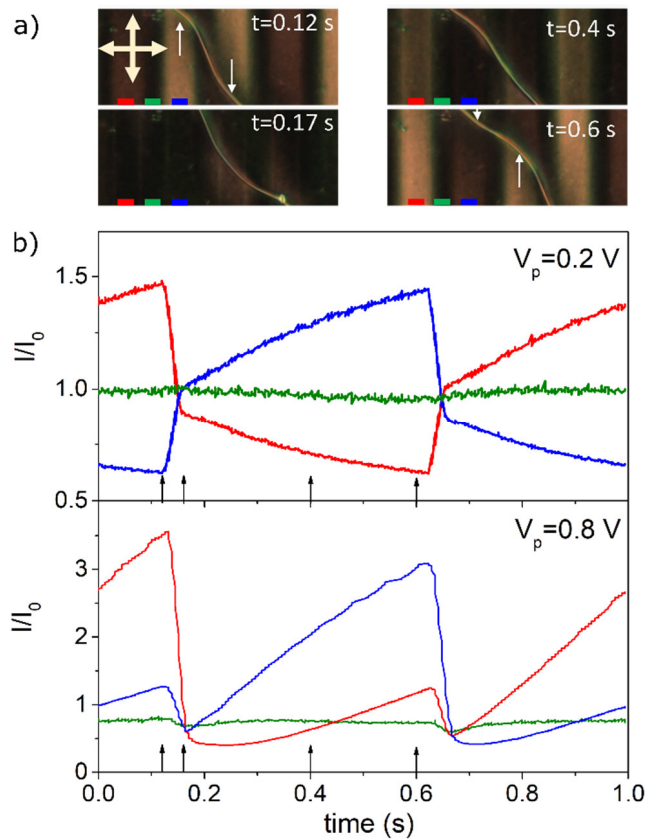

Fig. 7. (Color online) a) POM snapshots of the electro-optic response to the application of a square-wave (1Hz, $V_p$=0.8 V) at different times of the switching cycle for the rubbing direction of the sample cell at 0° with respect to the crossed polarizers. White arrows point towards the bend direction of the domain "soft" wall, indicating that the area with larger transmitted intensity under field expands, while that with lower transmitted intensity contracts. b) and c) Time dependence of the main intensity during a switching cycle for those areas marked in (a): red (one field direction), blue (opposite field direction) and green (on top of the electrode) for peak voltages of $V_p$=0.2 V and $V_p$=0.8 V. Black arrows correspond to the times of the snapshots in (a).

**Second Harmonic Generation**

To gain insight into the reorientation of the polarization direction under the application of in-plane perpendicular fields, we performed SHG imaging in the same IPS cells (see Methods). Measurements were performed with polarised incoming laser beam and no analyser. On cooling from the *N* phase, as already reported, a periodic stripped texture can be clearly observed by SHG imaging (Fig. 8.a), with a periodicity of around 10 μm (incoming polarization parallel to rubbing direction). On further cooling, the SHG texture becomes uniform and darkens after the propagation of the structure relaxation lines described previously (transition sequence shown in Fig.ESI.14). The dependence of the transmitted SHG intensity on the angle between rubbing direction and incoming polarization is shown in Fig. 8.b for a temperature after the structural relaxation. The observed angular dependence follows the dependency $I_{2\omega} \propto \cos^4 \phi$, indicating that the main SHG activity arises for the polarization of the pump laser along **n**.

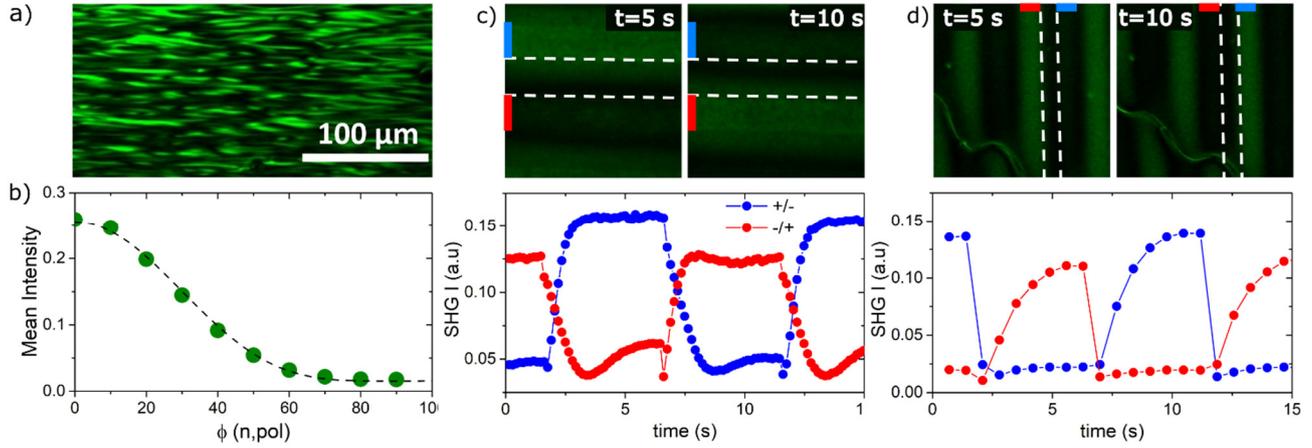

Fig. 8. (Color online) Second harmonic generation (SHG) a) Snapshot at the transition showing a periodic pattern with a periodicity of ~10 μm. b) Angle dependence of SHG signal, where φ is the angle between the incoming polarization and rubbing direction. Dashed line corresponds to the fit $\propto \cos^4 \phi$ c) SHG images of the switching response for a square wave of 0.1 Hz and Vp=0.75 V, when the initial **n** direction is parallel to the incoming light polarization. Bottom plot shows the time dependence of normalized SHG intensity during the switching. d) SHG images when the initial **n** direction is perpendicular to the incoming light polarization. Bottom plot shows the time dependence of normalized SHG intensity during the switching.

Due to small scanning rate of the confocal microscope, in plane switching was studied for 0.1 Hz square pulses for voltages of $V_p$= 0.75 V. The scanning rate of the confocal microscope was set to 1800 Hz resulting in an acquisition rate of 6.8 frames per second. Comparison of the wall deformation under fields allows correlating SHG results to those obtained by POM (Fig.ESI.14). When incoming polarization is parallel to rubbing, for the same polarity at which POM results show a decrease in the transmission intensity (for example red area in Fig. 7 at 0.5 seconds) the SHG signal initially significantly drops and then equilibrates to larger value, although lower than in the absence of electric field. For the opposite polarity (that at which transmitted intensity strongly increases in POM), SHG signal is larger and of the order of that observed without electric field. When incoming polarization is perpendicular to rubbing (Fig. 7.d, scan rate 700 Hz and 1.4 fps due to the larger scanned area), the SHG signal is very weak when the field is off in agreement with Fig. 8.b. On applying the field, SHG signal increases strongly for field polarity at which POM shows larger transmission (for example red area in Fig. 7 at 0.1 seconds, compare with wall deformation in Fig. 7.d) and almost no change can be detected for that polarity at which the POM transmission changes little.

**Discussion**

Comparison of the textural features observed in different types of commercial cells shows the strong effect that surfaces have on the final structure of the $N_S$ phase in confinement. While in those planar EHC cells with antiparallel rubbing (treated with polyimide LX-1400 from Hitachi-Kasei), only one type of domain has been observed, separated by "soft" walls, observations in planar Instec cells (KPI-300B) also with antiparallel rubbing have shown a broader hierarchy of structures. For the latter, even when comparing cells from the same batch (for example Instec IPS cells with rubbing antiparallel to the electrodes) big differences have been observed from cell to cell. For some of these cells, only big domains separated by "soft" walls are present while for others the whole area is covered by smaller domains divided with "hard" walls embedded in large domains of the previous type. Interestingly, as also observed in Instec IPS cells, although aligning agent and the antiparallel type of rubbing are the same, the *N-$N_S$* transition manifest very differently for cells with the rubbing parallel or perpendicular to the electrodes. The observed textural features depend not only on the surface properties, but also on the history of the sample/surface. It has been observed that the application of electric fields can result into the nucleation of "hard" wall-surrounded domains in areas where they were previously absent. Heating back to the $N$ phase and cooling down into the $N_S$ reveals that those "hard" walls are triggered by the local characteristic of the surface as they will reappear in the same areas. These observations indicate, not only that experimental observations for this new ferroelectric splay nematic phase have to be done by carefully controlling the confining media surfaces, but, most importantly, that the control of confining surfaces can be used to manipulate the macroscopic structure of the sample.

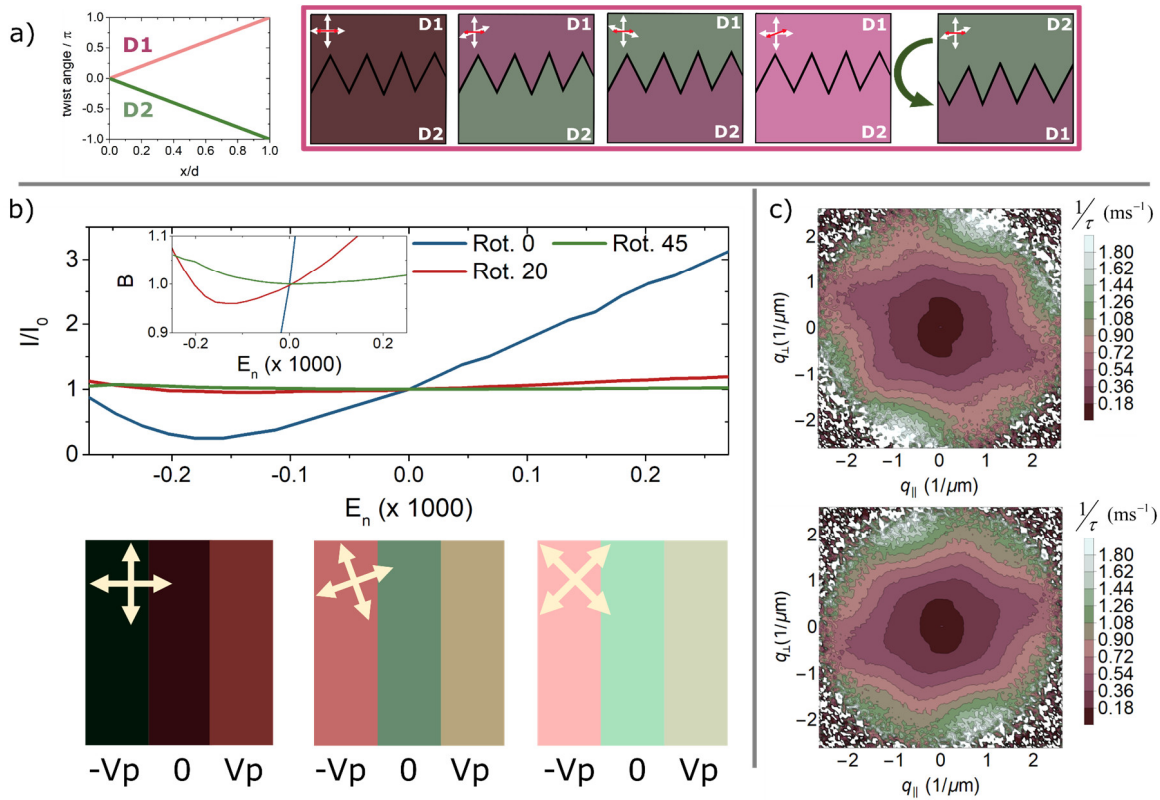

Fig. 9. (Color online) (a) Schematic of Berreman calculus for two domains with a full π-twist of opposite handedness. From left to right: twist angle in the LC cell, appearance under crossed polarizers with the cell rubbing direction aligned with the polarizers, uncrossing polarizers in opposite directions, rotating sample by 20 degrees with respect to crossed polarizers and sample upside-down flip along the rubbing direction. (b) Top: Calculated normalised transmitted intensity as a function of the normalized electric field for the structure given in (a). Bottom: Example of calculated optical behaviour for Vp=±0.0024. (c) q-dependence of relaxation rates for RM734 at 120 °C in the $N_S$ phase in two adjacent π-twist domains.

Optical observations summarized in Fig. 2 demonstrate that those domains predominantly found in our experiments exhibit twist deformations with opposite handedness. To gain insight into the structural twist features we performed Berreman calculations to describe the expected optical behaviour for different twist structures and experimental geometries. Taking into account a cell thickness of 8.3 μm and Δn=0.24 we investigated the resulting transmitted spectra for the three simplest twist geometries, i.e simple twist (Fig.ESI.15.a), double twist (Fig.ESI.15.b) and full-pi twist deformation (Fig. 9 and Fig.ESI.15.c for comparison), for the same conditions as experimentally shown in Fig. 2. Although the three twist structures can account for the optical symmetry revealed under uncrossing the analyser, a simple lens-like twist (Fig. SI 12 a) fails to account for the optical degeneracy observed under sample rotation with respect to the crossed polarisers. Double twist (Fig.ESI.15.b) and full-pi twist (Fig.ESI.15.c) structure can explain such observations. However, for the double twist structure, calculations show that when flipping upside-down the sample around the cell rubbing direction and with the analyser slightly uncrossed, domain optical activity is expected to reverse (Fig.ESI.15.b) which is not observed in the experiments. In short, only the full π-twist structure accounts for all the observations summarized in Fig. 2. Thus, left handed and right handed twist domains, are separated by "soft" walls, topologically 2π disclination lines, which are not pinned to the surface and can move. This is in agreement with those results presented in ref. 21, in which the appearance of the twist across the cell is attributed to the antiparallel unidirectional rubbing resulting in opposite in-plane polar anchoring in both surfaces. Charged surfaces result in a small electric field pointing out of the surface plane, which can force the sample polarization to orient out of the plane. In the presence of a cell pretilt, as is the case for antiparallel rubbing, pretilt and out of plane electric field will result into opposite orientational effect for the polarization at the cell surfaces, which can be the reason behind the formation of a full π-twist structure across the cell.[21] Probably it is interesting to recall at this point, the very different equilibrium states obtained for the $N_S$ phase in the Instec IPS cells (same aligning agent and antiparallel rubbing for both surfaces, i.e. pretilt) for the rubbing being parallel to be perpendicular to the electrodes. In the latter, no twist structural relaxation is observed resembling more the observations discussed for bidirectional rubbing with no pretilt.[21]

Additionally, results shown in Fig.ESI.15 can be related to the textural observations for those domains enclosed by the hard domains. In this case, rotation of the sample in opposite directions reveals different colour texture, indicating that the twist structure is different than the full π-twist. Comparison of experimental observations (Fig.ESI.3) and these calculations, immediately points to the presence of a simple twist structure as indicated in Fig.ESI.15. As mentioned before, such domains grow in areas where before the structural relaxation a lens-shape structure is observed, similar to those reported in reference 21 for surfaces with bidirectional rubbing. In Fig.ESI.3 it can be seen that, when one of these domains crosses several larger π-twist domains, texture colour alternates on sample rotation when going from one surrounding π-twist domain to the adjacent, indicating that handedness changes across the domain without giving rise to a wall (Fig.ESI.3 g-h). Additionally, when applying perpendicular in-plane fields, these domains show larger/smaller transmitted intensity for the field polarity at which the π-twist domain in which they are embedded shows larger/smaller transmission. These two observations together suggest that the simple twist structure is determined by the π-twist domain in which it is embedded and the position of the "hard" domain wall, which is pinned at one of the surfaces and is separating the regions having opposite polarization direction (Fig. ESI. 16). Returning to observations for IPS cells with rubbing perpendicular to the electrodes (Fig. ESI. 4, Fig. ESI. 5 and ESI. Video 3), comparison with Fig.ESI.15 would suggest the presence of a simple twist structure in these domains. Although weak, differences can be observed when uncrossing polarizers or rotating the sample, suggesting that the structure is not uniform and that the direction of the twist can change over long distances. Application of small voltages (Fig.ESI.5.f, in this case, field directed along the rubbing direction) shows that one of the field polarities is favourable for the domain polarization resulting in little reorientation, while the other polarity provokes a large reorientation of the material. Additionally, Fig.ESI.5.f evidences that the average polarization direction of neighbouring domains lay in opposite directions.

We used the same formalism to calculate the normalized transmitted intensities under the applications of in-plane fields directed perpendicularly to the cell rubbing direction as analysed in the electro-optic experiments presented here. For that, the previous required step is the calculation of the director structure under the application of an in-plane electric field. We considered a simple model in which the field reorients the polarization, which remains constant and parallel to the director. That being said, the application of an electric field will also increase the order of the dipoles and can also induce dipole moments, but such effects have not being taken into account for the shake of simplicity. In the absence of an applied field, the nematic director was considered to make a full uniform π-twist turn along the cell thickness. Equilibrium solutions under field where calculated as described in ESI. Section I. Considering a cell thickness of 9 μm and Δn=0.24, calculated normalized intensities are shown in Fig. 9.b as a function of a normalized applied field (ESI.Section I). Comparison with Fig. 6 shows good agreement with experimental findings, where for a given field polarity normalized transmitted intensity linearly grows at small fields while decreases for the other polarity reaching a minimum at sample rotations of 0 and 20 degrees. Even in the case of the sample rotated at 45 degrees with respect to the crossed polarizers calculations are in very good agreement with experimental results. Calculations not only accurately reproduce the experimental trend but also the observed values. One should keep in mind, that a more "complex" full-π twist, as in more localized structure, shows similar results than the continuous full-π twist, both for the observed characteristics under no field and for the field dependence of the transmitted intensity. This suggests that continuous π-twist is only one of the possible structures and that distinguishing between them requires experimental evidence beyond optical observation. The same Berreman calculations could be performed for SHG observations, however, one should bear in mind that such analysis is more complicated as it strongly depends on the details of the structure. As shown in Fig. 8.b the main contribution to the SHG signal arises from the polarization of the laser parallel to the cell rubbing direction. In theoretically calculating the SHG signal one should take into account the how well optical polarization is guided inside the sample, interference effects of the SHG beam and the effects arising from the increase of the polarization due to the application of electric fields. Although feasible, such calculations can lead to wrong conclusions if the primary requirement, the knowledge of the fine details of the structure, is missing.

To further characterize the structural twist we performed cross-differential dynamic microscopy (c-DDM, see Methods) measurements at 120 °C in two adjacent domains as those shown in Fig. 2 ($T_{N-Ns}$-$T$=12). c-DDM is a method equivalent to dynamic light scattering (DLS) (Fig. ESI.17), in which relaxation rates of thermal director fluctuations are measured. The advantage of the c-DDM over the DLS is that in a single measurement the relaxation rates at many values and directions of wavevector is measured. Here, we will only focus on the shape and the symmetry of the q-dependence of fluctuations, which reflects the underlying director structure. Deeper analysis and modelling of the results will be presented elsewhere. When the system is in the ordinary nematic phase, q-dependence surface plot is symmetric with respect to the rubbing, reflecting uniform orientation of the director, with faster increase of the relaxation rates with the wavevector parallel to **n** (see Fig. ESI. 16.b and c), due to faster bend fluctuations (Methods). When lowering the temperature in the *N* phase, slowing down of the fluctuations can be appreciated by comparison of q-dependence plots shown in Fig.ESI.16 b and c.[7] In the case of a uniform π-twist, it is expected that the relaxation rates do not depend on the direction of the wave vector, as is for example observed in a cholesteric liquid crystal (Fig. ESI.17 d and e). In the $N_S$ phase in the π-twist domains, the dispersion curves have star like shape (Fig. 9c), indicating that the structure is not a uniform twist, but the director orientation seems to have three preferred in-plane orientations separated by localized twist deformation. In order to gain more information about the structure in the main domain observed for IPS cells with the rubbing perpendicular to the electrodes, we also perform c-DDM measurements in those. Comparison of the results shown in Fig. ESI. 18 for the $N_S$ phase at 120 °C with those characteristic of the uniform *N* phase (Fig. ESI.17.b and c) shows that the main orientation of the director is tilted in plane with respect to the cell rubbing direction. Such results are compatible with the presence of a simple twist structure as also suggested by the textural features.

All these observations together open many questions. The appearance of twisted structures can be a consequence of the confinement and surface effects, however, it is also possible that the phase is intrinsically chiral. MD simulations showed that in the absence of the field the polarization is parallel to the average orientation of the molecules.[20] However, the polar order of molecular axis measured by $P_1 = \langle Cos(\theta) \rangle$ where $\theta$ is the angle between the director and the molecular axis is larger than the polar order calculated for the direction of the dipoles. That the two order parameters are different is expected because the molecular long axis and the dipole are not parallel (Fig. 1). While molecular axes are on average oriented along the director, the dipole moments are randomly oriented on a cone (Fig. ESI.19). It is expected, that an electric field will increase the order parameter of the dipoles, one way this can happen is that the dipoles preferentially orient at an angle on the cone determined by the direction of the field (Fig. ESI.19), resulting in a structure that is biaxial. This situation is similar but not equivalent to the biaxial ferromagnetic phase.[22] Such polar biaxial phase is determined by two vector order parameters, and has therefore different symmetry than the nonpolar biaxial phase which can be described by two tensors.[23] The response of the π-twist domains to an out-of-plane field (Fig. 3, Fig. ESI.6 and ESI.Video.4) depends on the handedness of the domain. At a field with a given sign, not only the domains with favourable handedness grow, but also new domains with this handedness emerge. One possible explanation of this behaviour is that the field in addition to polar biaxial order also induces chirality. In such case in an unconstrained sample, the structure in the homogeneous external field could resemble polar heliconical structure (Fig. ESI. 20) in which the pitch depends on the magnitude of the field.

Due to the flexoelectric coupling, polar phase exhibits instability towards splay deformation,[8] and, in the case that the director and polarization are parallel, the splay deformation would cause the divergence of polarization and consequently the depolarization field. This field can either suppresses the splay, or as discussed above, can induce biaxial order combined with chirality. The latter could explain, why the twist in π-twist domains is localized. The field at the surface causes twist in the regions at the surfaces, while inhomogeneous splay could cause it in the interior of the cell. Further testing of this hypothesis is above the scope of this paper.

All the experimental data reported here, demonstrate a wide variety of behaviours. Confinement, surfaces, surface charges, thermal history, field or field application history are just an example of all the parameters which affect the final behaviour of the sample. These lead to a hierarchy of polar structures which can be easily switched by electric fields, clearly showing polar linear coupling. Understanding, control and exploitation of all these behaviours open an exciting path for the study of this new phase. For instance, SHG results show just part of the big application potential of this new ferroelectric splay nematic phase to be employed in photonic applications going beyond the classical liquid crystals ones.

## Methods

**Polarizing optical microscope**

Polarizing optical microscopy experiments were performed in an Optiphot-2 POL Nikon microscope. Images and videos for domain formation and out of plane electric fields were recorded with a Canon EOS100D camera, while experiments in IPS Instec cells were recorded with a CMOS camera (BFLY-U3-23S6C-C). For IPS experiments, the repetition rate of the field profiles was set to 1 Hz and recorded for 400 frames at 15.9 fps. From the video, the response corresponding to a single cycle was then reconstructed. The sample was held in a heating stage (Instec HCS412W) together with a temperature controller (mK2000, Instec).

**Second harmonic generation Imaging**

Second harmonic generation imaging was performed in a using TCS SP8 Confocal Laser Scanning Microscope (Leica) equipped with a tunable IR laser and a polarising wheel. Experiments were performed in transmission with the laser excitation with 880 nm and acquisition 440/20 nm. Experiments were performed in transmission, where the incoming beam is stopped by a 400/20 nm filter before the detector. Experiments were performed with the incoming polarization in the horizontal direction of the shown images and no analyser was employed.

**Berreman calculus**

We computed the transmission spectra and color rendering using "dtmm" software package, which uses the Berreman 4x4 matrix method to calculate the transmission and reflection spectra. We first calculated the transmission spectra assuming standard daylight conditions using D65 standard illuminant. The calculated spectra were then converted first to XYZ color space using CIE 1931 color matching function. Next, we computed the linear RGB color from XYZ color space, as described in the sRGB standard IEC 61966-2-1:1999. Finally, we (optionally) applied the sRGB transfer function (gamma curve) to obtain the final nonlinear RGB color values suitable for display or print. In the absence of the sample and with uncrossed polarizers, the above-described procedure converts the input light source spectra to a neutral gray color. Therefore, to match the experimentally obtained images with the simulations, we performed in-camera white balance correction to get a neutral gray color of the microscope's light source. The white-balance correction performs matching of the camera's tri-stimulus values for a given light source to the ones obtained under daylight conditions, allowing us to have a good match between the simulated and experimentally obtained images.

Experiments done in the IPS cell were measured with a BFLY-U3-23S6C-C camera, with no gamma correction, and we left the data in the linear RGB color space. According to the manufacturer's specifications, the camera's IR filter has a short-pass filter characteristic with a cutoff at 620 nm. Therefore, we simulated the IR filter's short-pass characteristics for these experiments by limiting the light source spectra to the specified cutoff wavelength. We did not apply gamma correction to obtain results in Fig 9b. Experiments done in planar cells were done with a Canon EOS100D camera, which uses an sRGB transfer function. Therefore, we applied the gamma curve to the simulated linear RGB data to obtain the resulting images in Figs. 9a and ESI.15.

**Cross- Differential Dynamic Microscopy (c-DDM)**

The Cross-Differential Dynamic Microscopy method was setup as extensively described in ref. 19 by Arko and Petelin. The setup comprises an infinity-corrected microscope objective with a 20x magnification, a 20 mm focal length tube lens followed by a beam-splitter, and two identical cameras (Flir BFS-U3-04S2C-CS with 6.9 µm pixel size and 1/3 inch sensor). Translational and rotational mechanisms ensure precise alignment of their field of view, and we aligned the cameras using an alignment ruler. For illumination, we used the Koehler illumination setup using Thorlab's M565L3 mounted phosphor-converted LED with a nominal wavelength of 565 nm and a bandwidth of 100 nm. The LED was operated in a pulsed regime with a low duty cycle to prevent sample heating, driven with a current of 5 A. Cameras were triggered as described in ref. 19 to acquire two different sets of images with varying times of acquisition. Camera exposure time was 60 µs, and we set the shortest time delay between two consecutive frames on cameras to 256 µs. The image sequence is then cross Fourier analysed as a function of time delay using an open-source package "cddm" to obtain the normalized image cross-correlation function.[24]

Similarly to traditional dynamic light scattering experiments, c-DDM allows studying the orientational fluctuations of the director field. Liquid crystals are characterized by two fundamental modes, the splay-bend (β=1) and the twist-bend (β=2) modes with relaxation rates:

$$\frac{1}{\tau_\beta} = \frac{K_\beta q_\perp^2 + K_3 q_\parallel^2}{\eta_\beta(q)} \qquad (1)$$

Where $q_\parallel$ and $q_\perp$ are the components of the wave vector parallel and perpendicular to the director, $\eta_\beta$ are the viscosity coefficients and $K_i$ are the splay (i=1), twist (i=2) and bend (i=3) elastic constants. In the experiments, we aligned the polarizer and analyser with the planar-aligned director. For this geometry, one can show that if the wavevector $q$ of the analysis is along the director, then the obtained relaxation rate in the N phase corresponds to a pure bend-mode relaxation.[25] If the wave vector is orthogonal, the relaxation rate is that of a pure twist mode.[26]

**Quantum Chemical Calculations**

Computational chemistry was performed in Gaussian G09 rev D01[27] on the ARC3 machine at the University of Leeds. Calculations utilised the M06-HF hybrid DFT functional[29] and the aug-cc-pVTZ basis set.[30] The keywords Integral=UltraFine and SCF(maxcycles=1024) were used to ensure convergence, while a frequency calculation was used to confirm the absence of imaginary frequencies and so confirm the optimised geometries were true minima.

**Conflicts of interest**

There are no conflicts to declare.

**Acknowledgements**

N. S., A. P. and A. M. acknowledge the financial support from the Slovenian Research Agency (research core Funding No. P1-0192, and research grants No. J7-9399 and No. J1-2459). RJM acknowledges the use of ARC3, part of the High Performance Computing facilities at the University of Leeds. AE acknowledges the support by Deutsche Forschungsgemeinschaft (DFG), Projects ER 467/8-2 and ER 467/17-1, DFG Major Research Instrumentation Program Project 329479947.

**References**

1    M. Born, *Sitzungsber Preuss Akad Wiss*, 1916, **30**, 614.
2    P. Palffy-Muhoray, M. A. Lee and R. G. Petschek, *Phys. Rev. Lett.*, 1988, **60**, 2303–2306.
3    F. Bisi, A. M. Sonnet and E. G. Virga, *Phys. Rev. E*, 2010, **82**, 041709.
4    S. Dhakal and J. V. Selinger, *Phys. Rev. E*, 2010, **81**, 031704.
5    R. Berardi, M. Ricci and C. Zannoni, *Ferroelectrics*, 2004, **309**, 3–13.
6    P. De Gregorio, E. Frezza, C. Greco and A. Ferrarini, *Soft Matter*, 2016, **12**, 5188–5198.
7    A. Mertelj, L. Cmok, N. Sebastián, R. J. Mandle, R. R. Parker, A. C. Whitwood, J. W. Goodby and M. Čopič, *Phys. Rev. X*, 2018, **8**, 041025.
8    N. Sebastián, L. Cmok, R. J. Mandle, M. R. de la Fuente, I. Drevenšek Olenik, M. Čopič and A. Mertelj, *Phys. Rev. Lett.*, 2020, **124**, 037801.
9    R. J. Mandle and A. Mertelj, *Phys. Chem. Chem. Phys.*, 2019, **21**, 18769–18772.

10  R. J. Mandle, S. J. Cowling and J. W. Goodby, *Chem. - Eur. J.*, 2017, **23**, 14554–14562.
11  R. J. Mandle, S. J. Cowling and J. W. Goodby, *Phys. Chem. Chem. Phys.*, 2017, **19**, 11429–11435.
12  P. L. M. Connor and R. J. Mandle, *Soft Matter*, 2020, **16**, 324–329.
13  X. Chen, E. Korblova, D. Dong, X. Wei, R. Shao, L. Radzihovsky, M. A. Glaser, J. E. Maclennan, D. Bedrov, D. M. Walba and N. A. Clark, *Proc. Natl. Acad. Sci.*, 2020, 202002290.
14  H. Nishikawa, K. Shiroshita, H. Higuchi, Y. Okumura, Y. Haseba, S. Yamamoto, K. Sago and H. Kikuchi, *Adv. Mater.*, 2017, **29**, 1702354.
15  J. Li, H. Nishikawa, J. Kougo, J. Zhou, S. Dai, W. Tang, X. Zhao, Y. Hisai, M. Huang and S. Aya, *ArXiv201114099 Cond-Mat Physicsphysics*.
16  M. Čopič and A. Mertelj, *Phys. Rev. E*, 2020, **101**, 022704.
17  S. M. Shamid, S. Dhakal and J. V. Selinger, *Phys Rev E*, 2013, **87**, 52503.
18  M. P. Rosseto and J. V. Selinger, *ArXiv200312893 Cond-Mat*.
19  M. Arko and A. Petelin, *Soft Matter*, 2019, **15**, 2791–2797.
20  R. J. Mandle, N. Sebastián, J. Martinez-Perdiguero and A. Mertelj, *ArXiv201102722 Cond-Mat*.
21  X. Chen, E. Korblova, M. A. Glaser, J. E. Maclennan, D. M. Walba and N. A. Clark, *ArXiv201215335 Cond-Mat*.
22  Q. Liu, P. J. Ackerman, T. C. Lubensky and I. I. Smalyukh, *Proc. Natl. Acad. Sci.*, 2016, **113**, 10479–10484.
23  R. Rosso, *Liq. Cryst.*, 2007, **34**, 737–748.
24  Andrej Petelin, *Cross-differential dynamic microscopy v. 0.2*, http://doi.org/10.5281/zenodo.3800382, 2020.
25  F. Giavazzi, S. Crotti, A. Speciale, F. Serra, G. Zanchetta, V. Trappe, M. Buscaglia, T. Bellini and R. Cerbino, *Soft Matter*, 2014, **10**, 3938.
26  A. Mertelj, N. Osterman, D. Lisjak and M. Copič, *Soft Matter*, 2014, **10**, 9065–72.
27  M. J. Frisch, G. W. Trucks, H. B. Schlegel, G. E. Scuseria, M. A. Robb, J. R. Cheeseman, G. Scalmani, V. Barone, G. A. Petersson, H. Nakatsuji, X. Li, M. Caricato, A. Marenich, J. Bloino, B. G. Janesko, R. Gomperts, B. Mennucci, H. P. Hratchian, J. V. Ortiz, A. F. Izmaylov, J. L. Sonnenberg, D. Williams-Young, F. Ding, F. Lipparini, F. Egidi, J. Goings, B. Peng, A. Petrone, T. Henderson, D. Ranasinghe, V. G. Zakrzewski, J. Gao, N. Rega, G. Zheng, W. Liang, M. Hada, M. Ehara, K. Toyota, R. Fukuda, J. Hasegawa, M. Ishida, T. Nakajima, Y. Honda, O. Kitao, H. Nakai, T. Vreven, K. Throssell, J. A. Montgomery, Jr., J. E. Peralta, F. Ogliaro, M. Bearpark, J. J. Heyd, E. Brothers, K. N. Kudin, V. N. Staroverov, T. Keith, R. Kobayashi, J. Normand, K. Raghavachari, A. Rendell, J. C. Burant, S. S. Iyengar, J. Tomasi, M. Cossi, J. M. Millam, M. Klene, C. Adamo, R. Cammi, J. W. Ochterski, R. L. Martin, K. Morokuma, O. Farkas, J. B. Foresman, and D. J. Fox., *Gaussian 09, Revision A.02*, Gaussian, Inc., 2016.
28  Y. Zhao and D. G. Truhlar, *J. Phys. Chem. A*, 2006, **110**, 13126–13130.
29  S. Grimme, J. Antony, S. Ehrlich and H. Krieg, *J. Chem. Phys.*, 2010, **132**, 154104.
30  R. A. Kendall, T. H. Dunning and R. J. Harrison, *J. Chem. Phys.*, 1992, **96**, 6796–6806.

# Electronic Supporting Information

# Electrooptics of mm-scale polar domains in the ferroelectric splay nematic phase

Nerea Sebastián,[1*] Richard J. Mandle,[2] Andrej Petelin,[1,3] Alexey Eremin[4] and Alenka Mertelj[1*]

[1]Jožef Stefan Institute, P.O.B 3000, SI-1000 Ljubljana, Slovenia

[2] School of Physics and Astronomy, University of Leeds, Leeds, UK, LS2 9JT

[3]Faculty of Mathematics and Physics, University of Ljubljana, Slovenia

[4]Department of Nonlinear Phenomena, Institute for Experimental Physics Otto von Guericke University Magdeburg Universitätsplatz 2, 39106 Magdeburg (Germany)

# Contents

A) Domain formation at different thicknesses in EHC cells

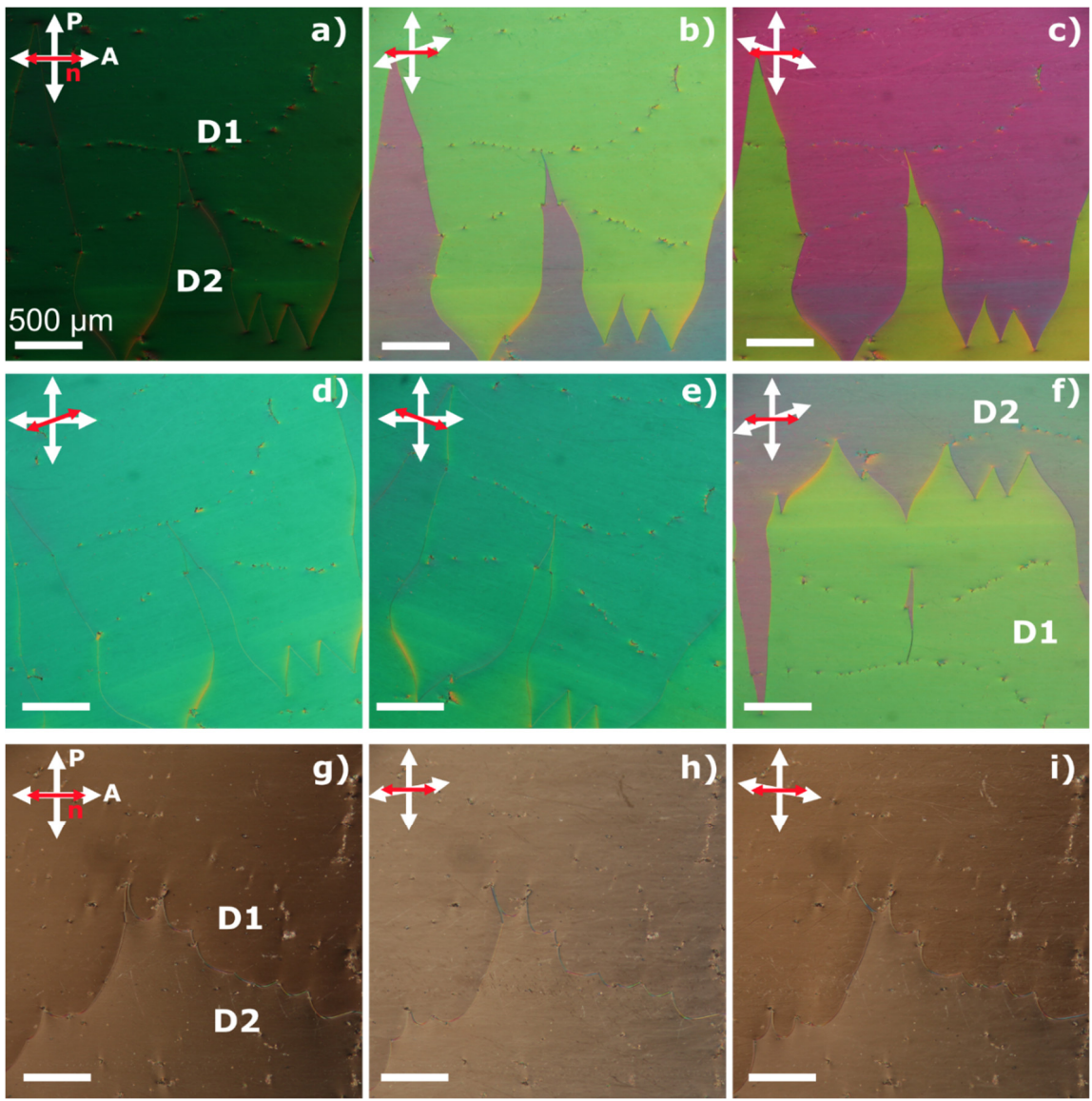

Fig. ESI.1 (Color online). Domains 5 and 25 microns EHC cells. (a-f) POM images of domains in 5.4 microns cell for (a) cell rubbing direction parallel to polarizers, (b-c) analyser uncrossed in opposite directions, (d-e) sample rotated in opposite direction between crossed polarisers and (f) sample flipped up-side-down and analyser uncrossed in the same direction as in (b). (g-i) POM images of domains in 22 microns cell for (a) cell rubbing direction parallel to polarizers and (b-c) analyser uncrossed in opposite directions.

## B) Formation and investigations of domains surrounded by "hard" walls

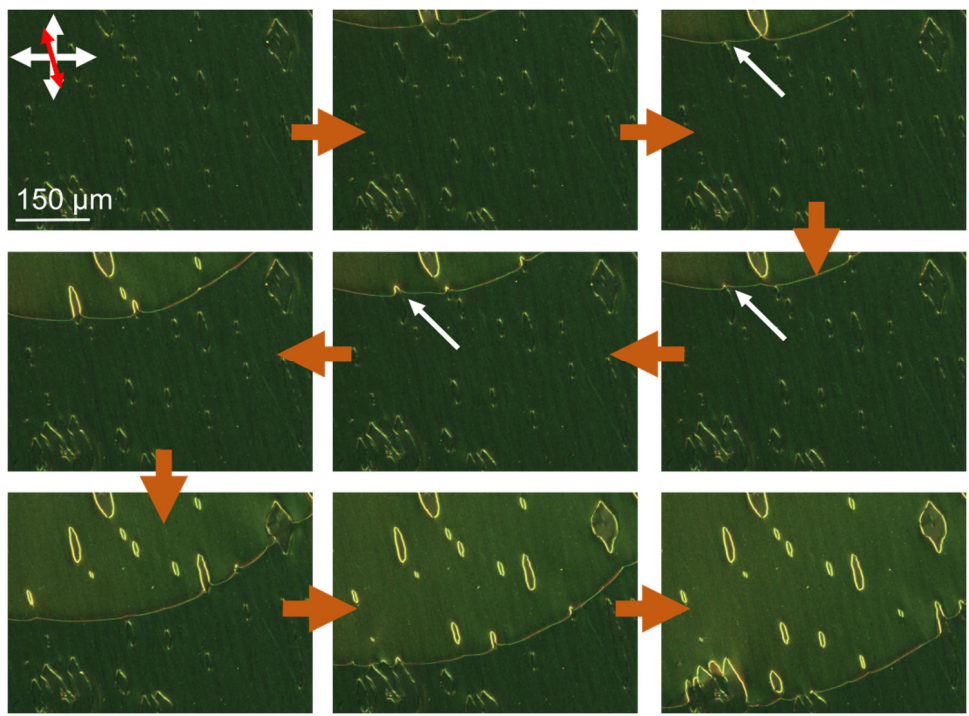

Fig. ESI. 2 (Color online) Formation of the "hard" wall domains during the structural relaxation. Orange arrows indicate the temperature sequence order starting from the sample in the Ns phase before the structural relaxation where some pre-existing lens-type structures are present. Propagation of the π-wall occurs smoothly across the sample. When the π -wall reaches of the lens-shaped structures, it splits and surrounds it without traveling through it, giving rise to a surface-pinned bright wall.

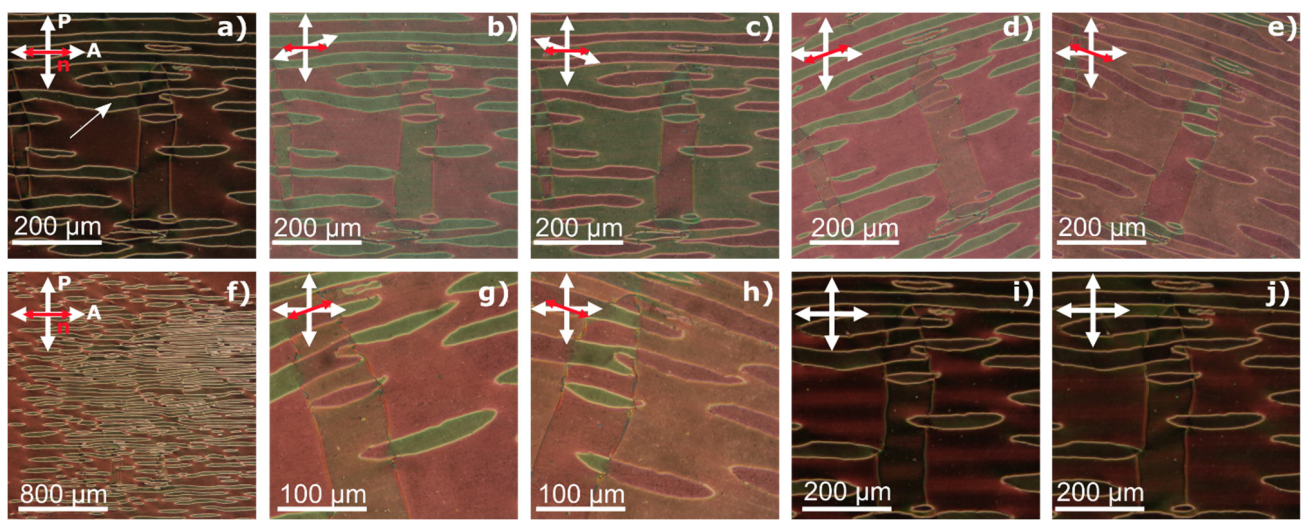

Fig. ESI. 3 (Color online) "Hard" domains. Observation of domains delimited by "hard" walls (example pointed by white arrow image a) in a background of π-twist domains. a) Extinction, (b-c) uncrossing analyser in opposite directions and (d-e) rotating the sample with respect to extinction position between crossed polarizers in opposite directions. While isolated "hard" domains can be found, it is common to find large areas covered by them (f). (g-f) Detail of the optical behaviour when rotating the sample. Areas within the hard domains show contrasting behaviour for opposite rotation. It is interesting to note that when one of such domains "crosses" a 2π- wall the behaviour is reversed. (i-j) Snapshots of texture observed under the application of 1Hz square-wave and $V_p$=0.5 V in the same area.

C) Transition in Instec IPS cells with cell rubbing perpendicular to the electrodes

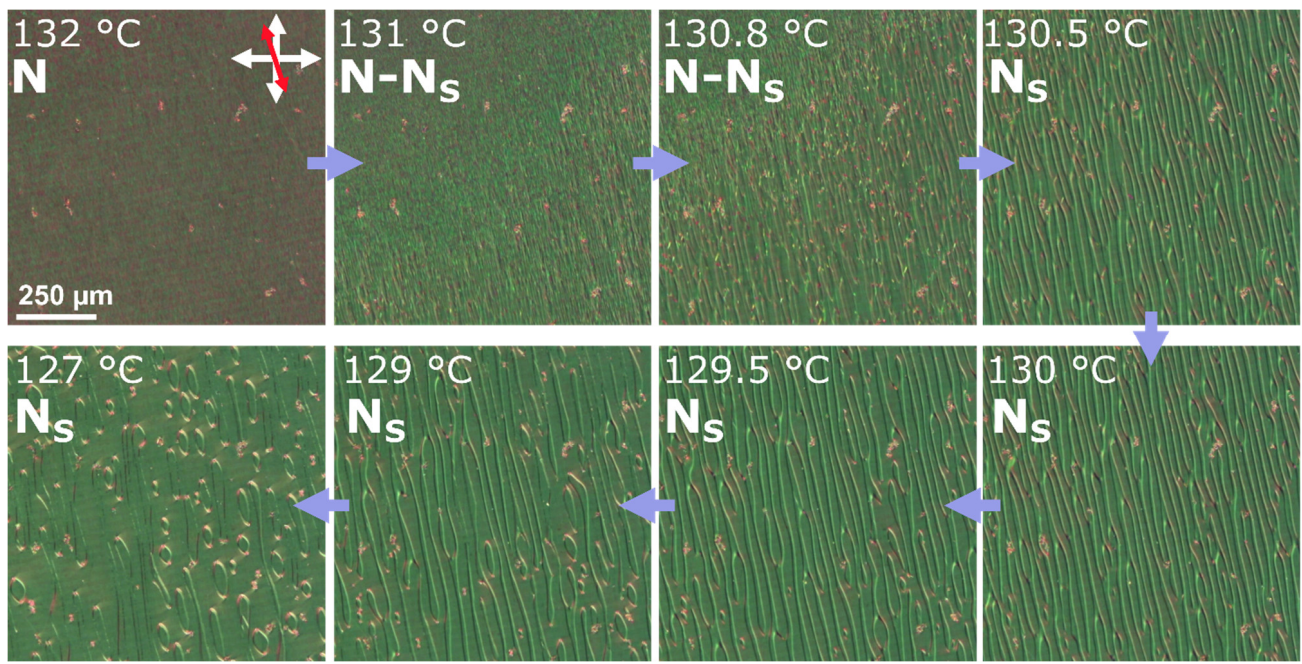

Fig. ESI.4. (Color online) $N$-$N_S$ transition on Instec IPS cells with cell rubbing perpendicular to the electrodes. Sequence shows the formation of elongated structures right after the stripped texture that slowly evolve into lens-shaped structures. No π-wall structural relaxation was observed in this case traveling the sample. In ESI Video 3 structural relaxation can be observed in the walls between domains, resulting in the walls observed in Fig. ESI.5.

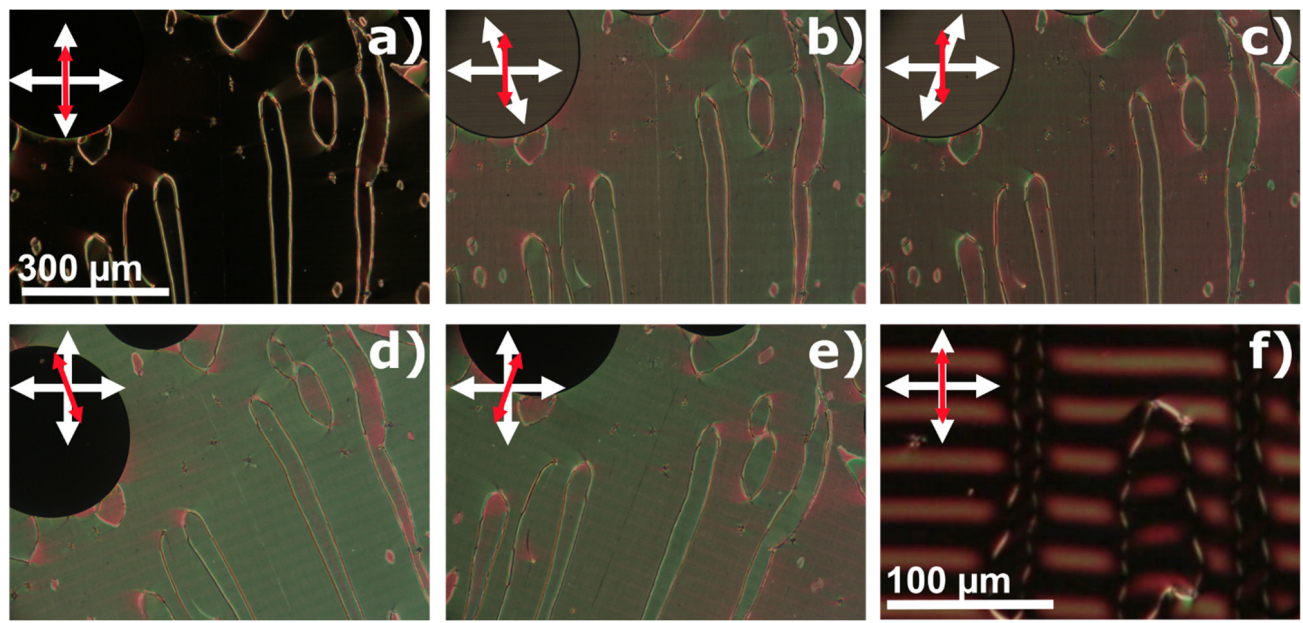

Fig. ESI.5 (Color online) Textural features at 120 °C in an Instec IPS cell with the rubbing direction perpendicular to the electrodes. (a) Extinction, (b-c) uncrossing analyser in opposite directions and (d-e) rotating the sample with respect to extinction position between crossed polarizers in opposite directions. (f) Snapshot of application of square-wave 1 Hz in-plane electric field along the rubbing direction showing opposite electro-optic behaviour from one domain to the other.

## D) Out of plane fields

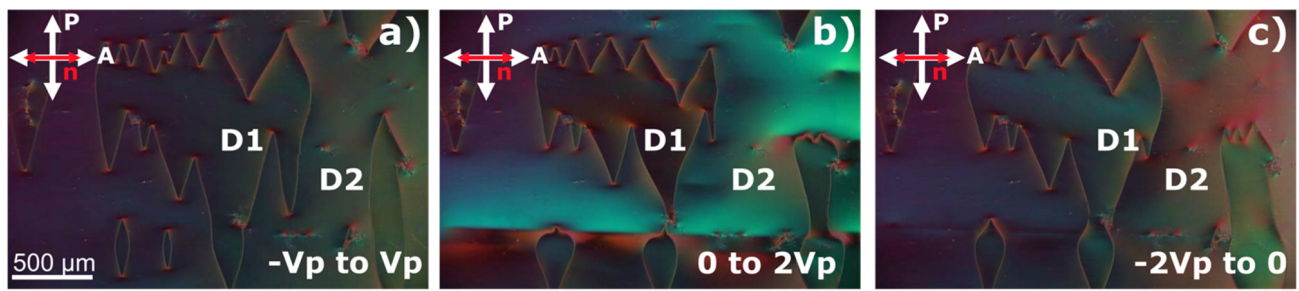

Fig. ESI.6 (Color online) Observation of a domain's size change under application of out of plane square wave electric field of Vp=0.5 V and 1 Hz in an EHC 5.4 µm cell. (a) For voltages alternating between –Vp and Vp domain size remains unchanged. (b) When an offset is applied and voltage alternates between 0 V and 2 Vp, the domain in the centre of the image progressively shrinks. (c) Finally, when polarity is reversed and voltage alternates between -2 Vp and 0, the studied domain progressively grows again.

## E) In-plane fields: short-pulses

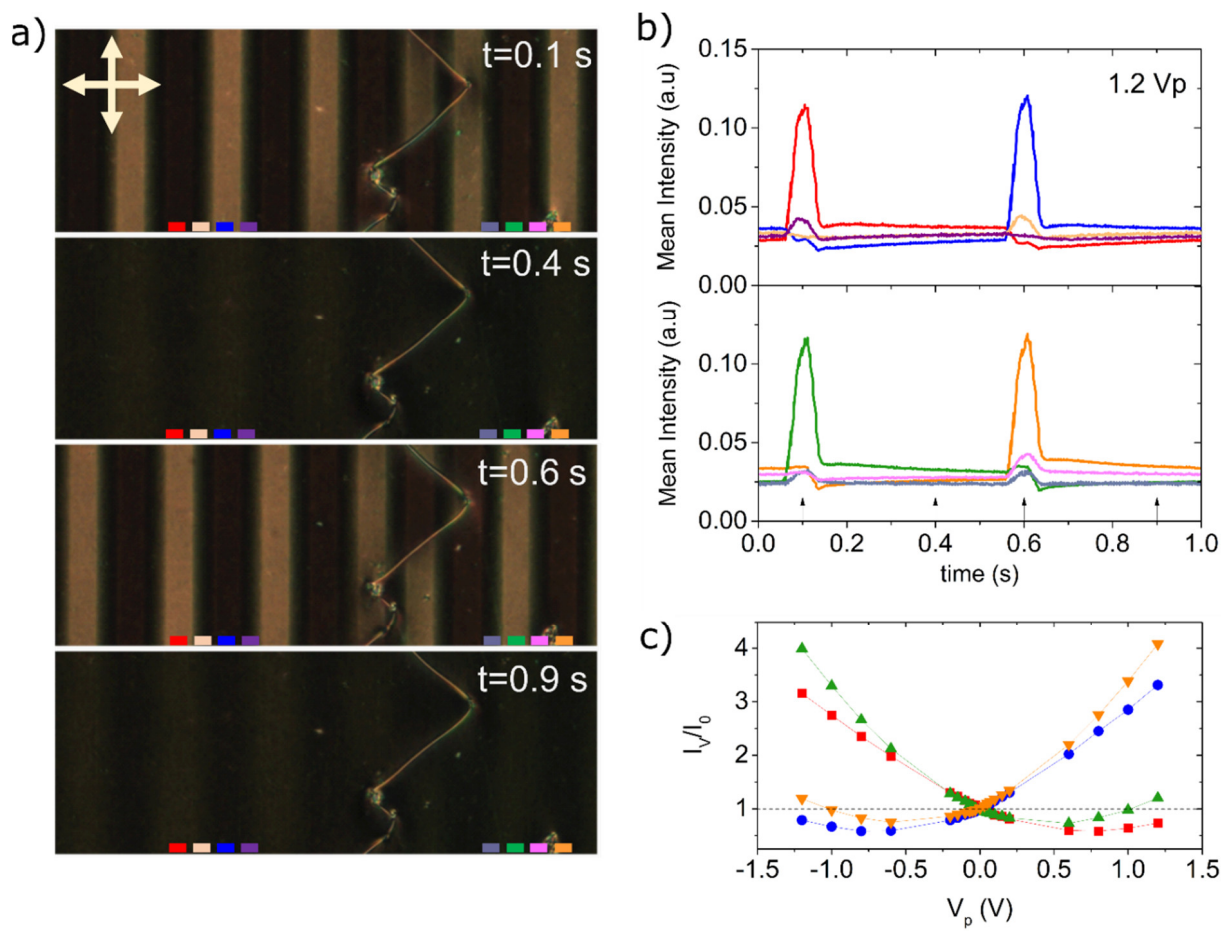

Fig. ESI.7. (Color online) Comparison between the electro-optic responses of both domains for the sample oriented with rubbing directions at 0 degrees with respect to crossed polarizers. a) POM snapshots of the switching sequence at four different times corresponding to the peaks of the pulses and 0 field plateaus. b) Time dependence over the switching cycle of the transmitted intensity, for a given $V_p$ for the areas marked in a). Left domain: red (one field direction), blue (opposite field direction), light-orange (over one electrode) and purple (over the next electrode). Right domain: green (one field direction, same as blue), orange (opposite field direction, same as red), grey (over one electrode, same as light-orange) and pink (over the next electrode, same as purple). Time has been shifted 0.1 seconds for better visualization. c) Normalized peak intensity vs pulse voltage for the four areas between electrodes.

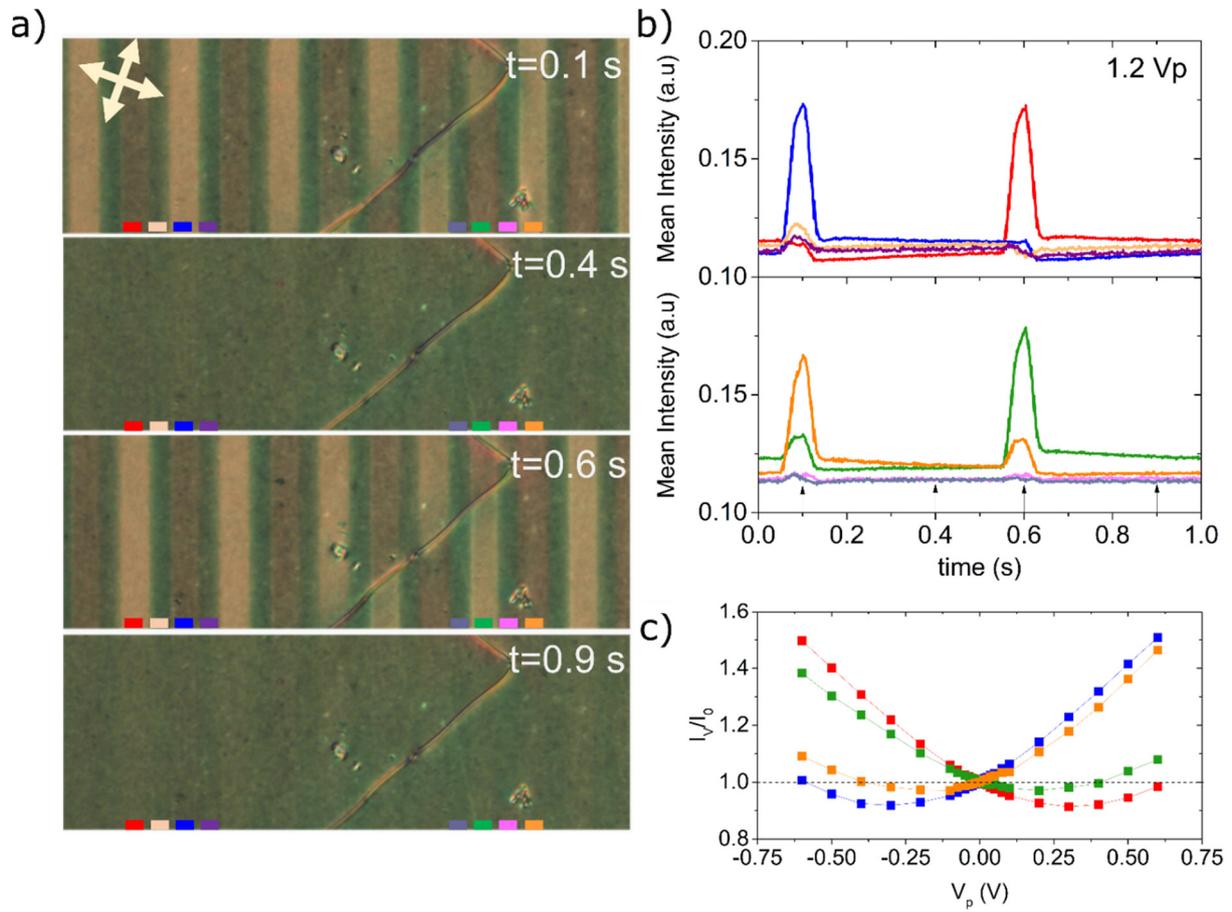

Fig. ESI.8. (Color online) Comparison between the electro-optic responses of both domains for the sample oriented with rubbing directions at 20 degrees with respect to crossed polarizers. a) POM snapshots of the switching sequence at four different times corresponding to peaks of the pulses and 0 field plateaus. b) Time dependence over the switching cycle of the transmitted intensity, for a given $V_p$ for the areas marked in a). Left domain: red (one field direction), blue (opposite field direction), light-orange (over one electrode) and purple (over the next electrode). Right domain: green (one field direction, same as blue), orange (opposite field direction, same as red), grey (over one electrode, same as light-orange) and pink (over the next electrode, same as purple). Time has been shifted 0.1 seconds for better visualization. c) Normalized peak intensity vs pulse voltage for the four areas between electrodes.

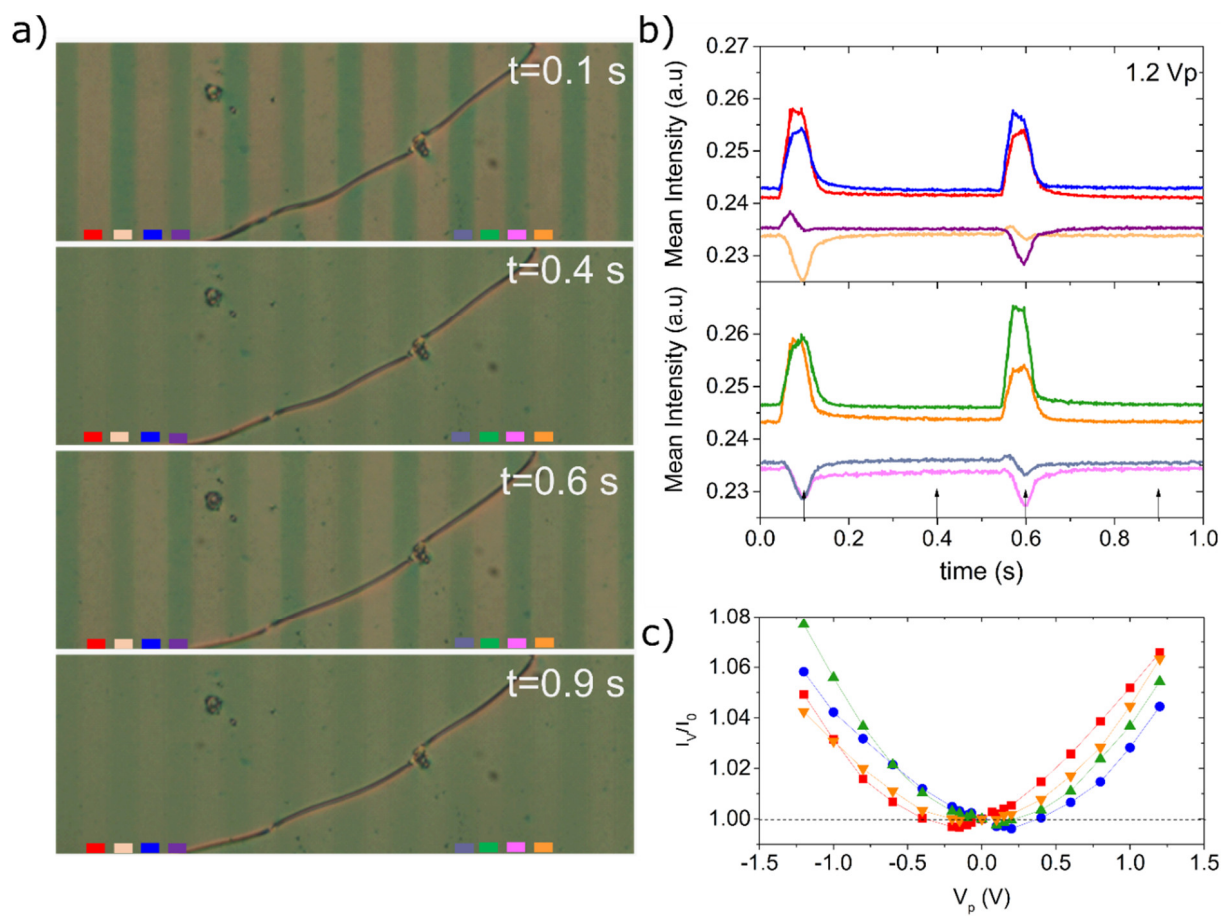

Fig. ESI.9. (Color online) Comparison between the electro-optic responses of both domains for the sample oriented with rubbing directions at 45 degrees with respect to crossed polarizers. a) POM snapshots of the switching sequence at four different times corresponding to the peaks of the pulses and 0 field plateaus. b) Time dependence over the switching cycle of the transmitted intensity, for a given $V_p$ for the areas marked in a). Left domain: red (one field direction), blue (opposite field direction), light-orange (over one electrode) and purple (over the next electrode). Right domain: green (one field direction, same as blue), orange (opposite field direction, same as red), grey (over one electrode, same as light-orange) and pink (over the next electrode, same as purple). Time has been shifted 0.1 seconds for better visualization. c) Normalized peak intensity vs pulse voltage for the four areas between electrodes.

F) In-plane fields: square-wave

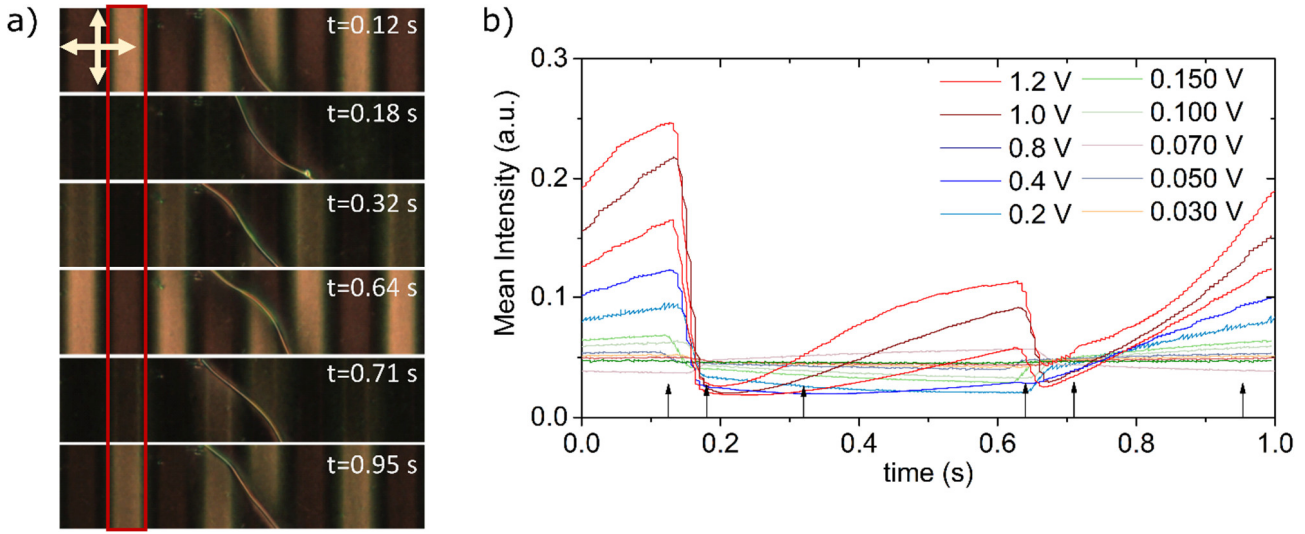

Fig. ESI.10 (Color online) a) POM snapshots of the electro-optic response to the application of a square-wave (1Hz, $V_p$=0.8 V) at different times of the switching cycle for the rubbing direction of the sample cell at 0° with respect to the crossed polarizers. b) Time dependence of the mean intensity over a switching cycle for different peak voltages for the highlighted area in (a). Arrows indicate times corresponding to the snapshots in (a).

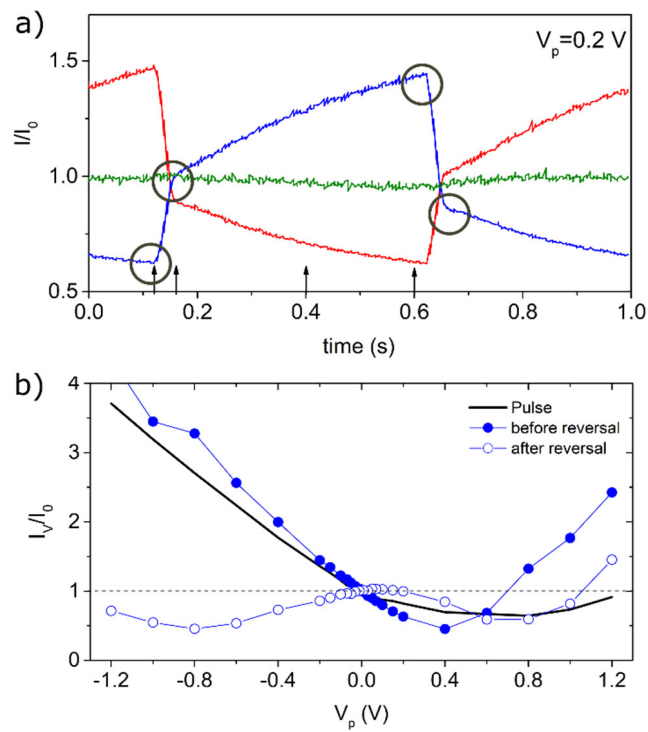

Fig. ESI.11 (Color online) a) Time dependence of the main intensity during a switching cycle for those areas marked in Fig. 6: red (one field direction), blue (opposite field direction) and green (on top of the electrode) for peak voltages of $V_p$=0.2. b) Comparison of the normalized mean intensity $I_V / I_0$ over reorientation for the same field area within a domain for short pulses (line), corresponding to the time before field reversal (full symbols) and corresponding time after field reversal (empty symbols). Such times are indicated in (a) by grey circles.

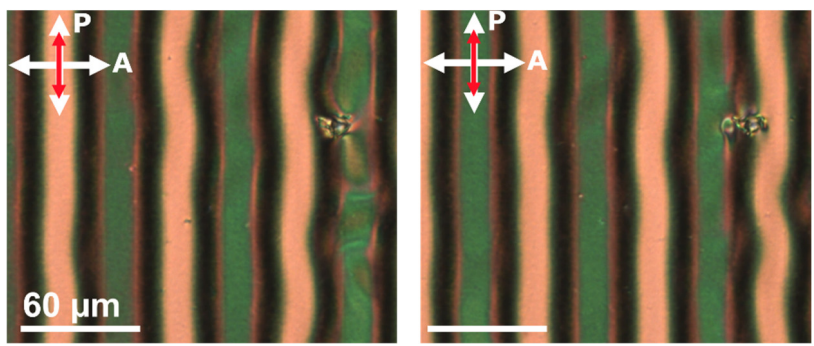

Fig. ESI. 12. (Color online) Snapshots taken during application of 1 Hz square-wave with Vp=2.8 V at opposite polarities showing the onset of the structure destabilization.

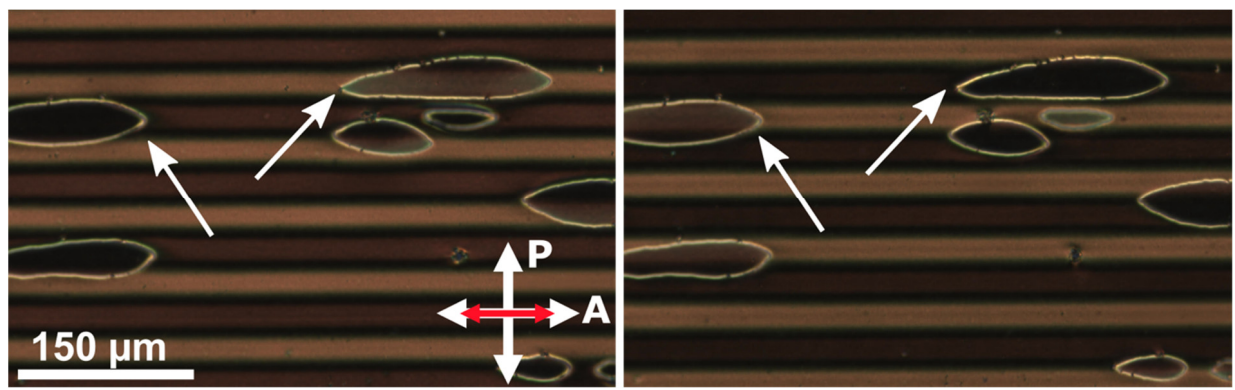

Fig. ESI.13. (Color online) Snapshots taken during application of 1Hz square-wave with Vp=1 V at opposite polarities in an area with several domains enclosed by "hard"-walls. Images reveal that such domains show increased/decreased transmitted intensity at the same field direction that the π-twist domain they are embedded in.

G) Second Harmonic Generation

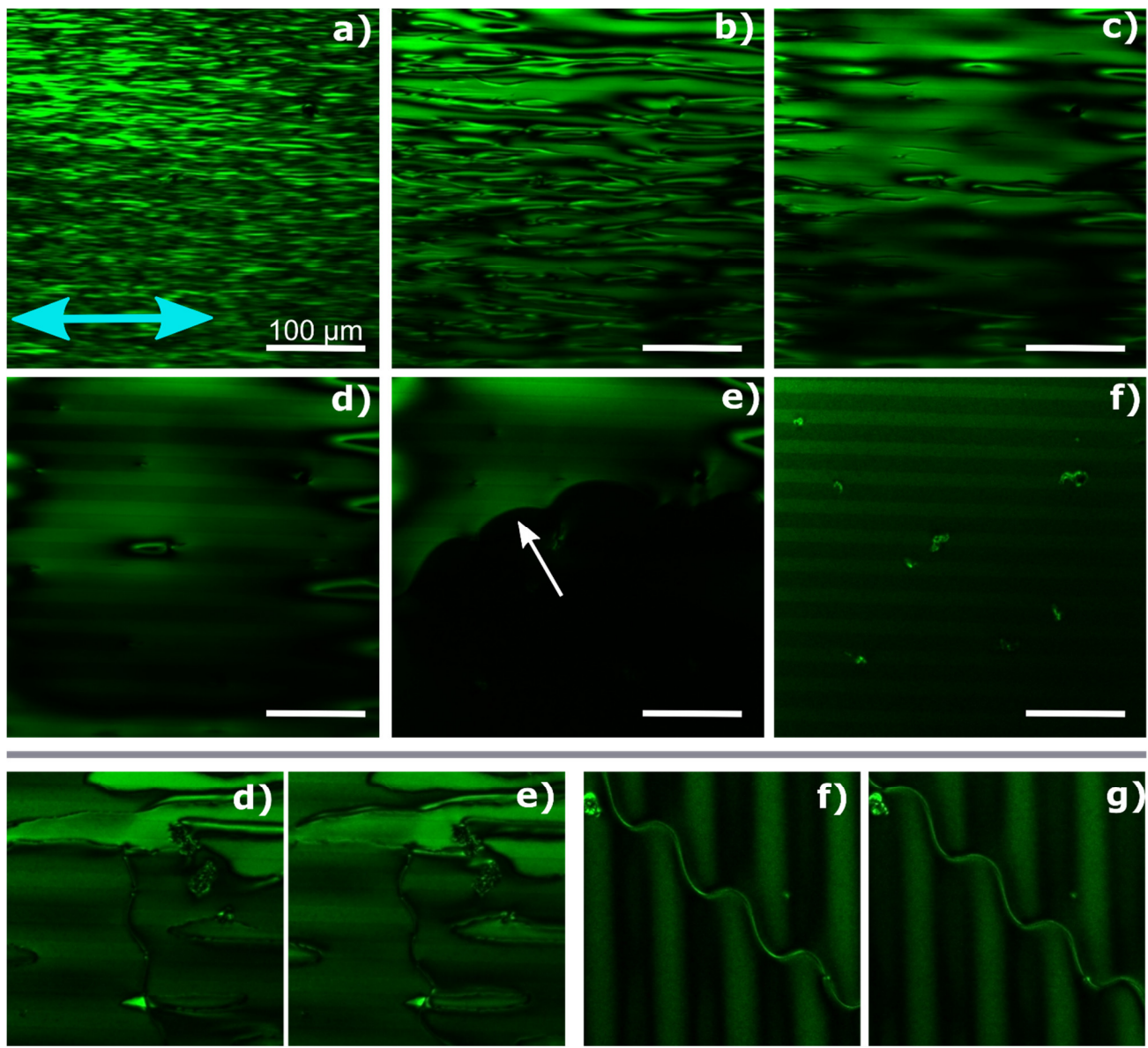

Fig. ESI. 14. (Color online). Second harmonic generation (SHG). The blue arrow indicates the polarization direction of the incoming light. No analyser is used. (a-f) SHG snapshots during N-Ns transition corresponding to the sequence also reported in Fig. 2. (a-d) in the main manuscript. Namely: a) destabilization of the homogeneous director (b-d) recovery of the homogeneous director, (e) propagation of π-wall across the sample and (f) SHG signal in the final domains state. Image (f) was taking by increasing gain x1.3 with respect to images a to f. Pictures (d-g) show different snapshots under application of electric field, square wave 0.1 Hz and Vp=0.75 V. (d-e) Two opposite field polarities and incoming light polarization parallel to the rubbing direction. (f-g) Two opposite field polarities and incoming light polarization perpendicular to the rubbing direction. Comparison of wall deformation allows correlating POM and SHG experiments.

## H) Berreman calculus for different structures

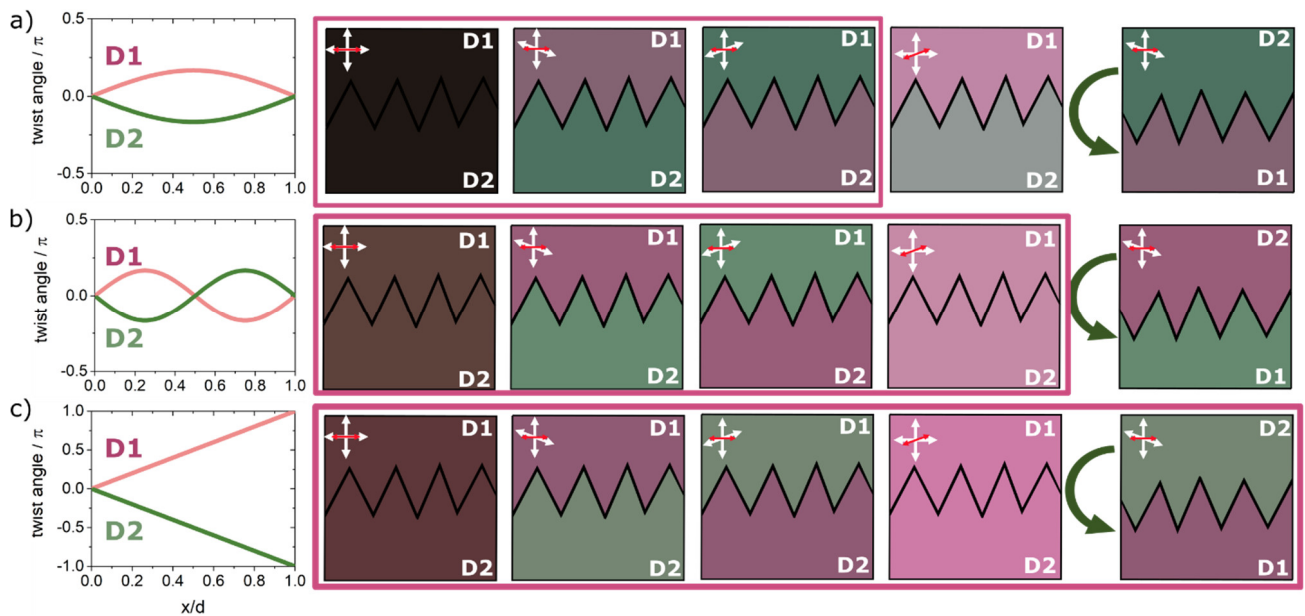

Fig. ESI.15 (Color online). Berreman calculus for different twist structures. (a) Domains with opposite handedness single twist, (b) Domains with opposite handedness double twist and (c) domains with opposite handedness π-twist. Left plots show the twist structure used in the calculations, while in the right results are shown for (from left to right) domains in extinction position, uncrossing analyser in one direction, uncrossing analyser in the opposite direction, rotating the sample in one direction and finally, upside-down flipping of the sample around the cell rubbing direction. Enclosed in rectangles are those results that match with the experimentally observed behaviour shown in Fig. 2 of the main manuscript.

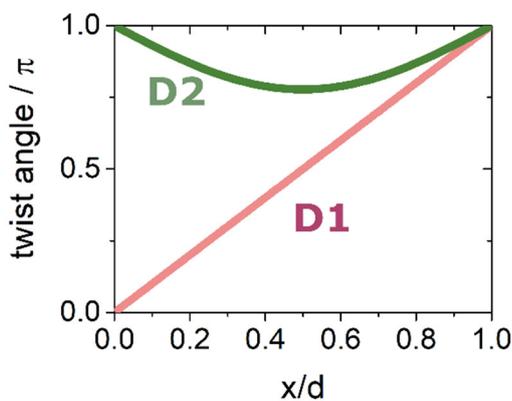

Fig. ESI.16 (Color online) Comparison of twist structures for simple twist and π-twist. Optically, simple twist depicted here is equivalent to that of Fig. ESI.17 for D2. For these two structures polarization at one surface is in the same direction while twisted by π in the other cell surface.

I) Dynamic calculation

Assuming an equilibrium twist structure, the free energy density in the $N_S$ phase was written as:

$$f = \frac{1}{2}K_2(\mathbf{n} \cdot \nabla \times \mathbf{n} - \phi'_0(x))^2 - \mathbf{P} \cdot \mathbf{E}$$

Where $K_2$ is the twist elastic constant, $\mathbf{n}=(\cos(\phi),\sin(\phi),0)$, the polarization is considered parallel to the director $\mathbf{P} = P_0\mathbf{n}$ and $\phi'_0(x)$ has the same role as the pitch in cholesteric liquid crystals, but in this case is dependent on the position along the cell in order to account for the initial twist structure. In the case of the π-twist structure considered for the calculation of the normalized transmitted intensity ($I/I_0$) we take a uniform π-twist $\phi_0(x)= \pi x$. The second term describe the coupling of the electric field with the polarization, which magnitude is assumed to be constant.

The final equilibrium structure under the application of electric fields was calculated by solving the dynamic equation:

$$\frac{\partial \phi(x,t)}{\partial t} = \frac{1}{\eta_\mathbf{n}}\left(\frac{\partial f}{\partial \phi(x,t)} - \frac{\partial}{\partial x}\frac{\partial f}{\partial \phi(x,t)/\partial x}\right)$$

Where $\eta_\mathbf{n}$ is the director rotational viscosity. The effect of anchoring on the surfaces was taken into account by considering

$$W\phi(x=0,t) = \partial \phi(x=0,t)/\partial x$$

$$W\phi(x=d,t) = \partial \phi(x=d,t)/\partial x$$

where W is the strength of the surface anchoring (in the present case considered to be strong W= 2·10⁻⁴ J/m²) and d is the cell thickness. We used P=5 µC/cm² and $\eta_n$=0.1Pa s (ref. 13). Normalized electric fields ($E_n = E_0 d/\sqrt{K_2\varepsilon_0}$) given in Fig. 9 in the main manuscript span from -0.00027 to 0.00027. One should take into account here, that the presented model is simplified, in which it is neglected that the internal field is inhomogeneous in the plane of the sample and is changing during the reorientation of the polarization. Given the polarization values of the material such effects would have a great impact in terms of local fields. Nominal experimental applied fields are of the same order as those employed in reference 13.

J) cross-Differential Dynamic Microscopy

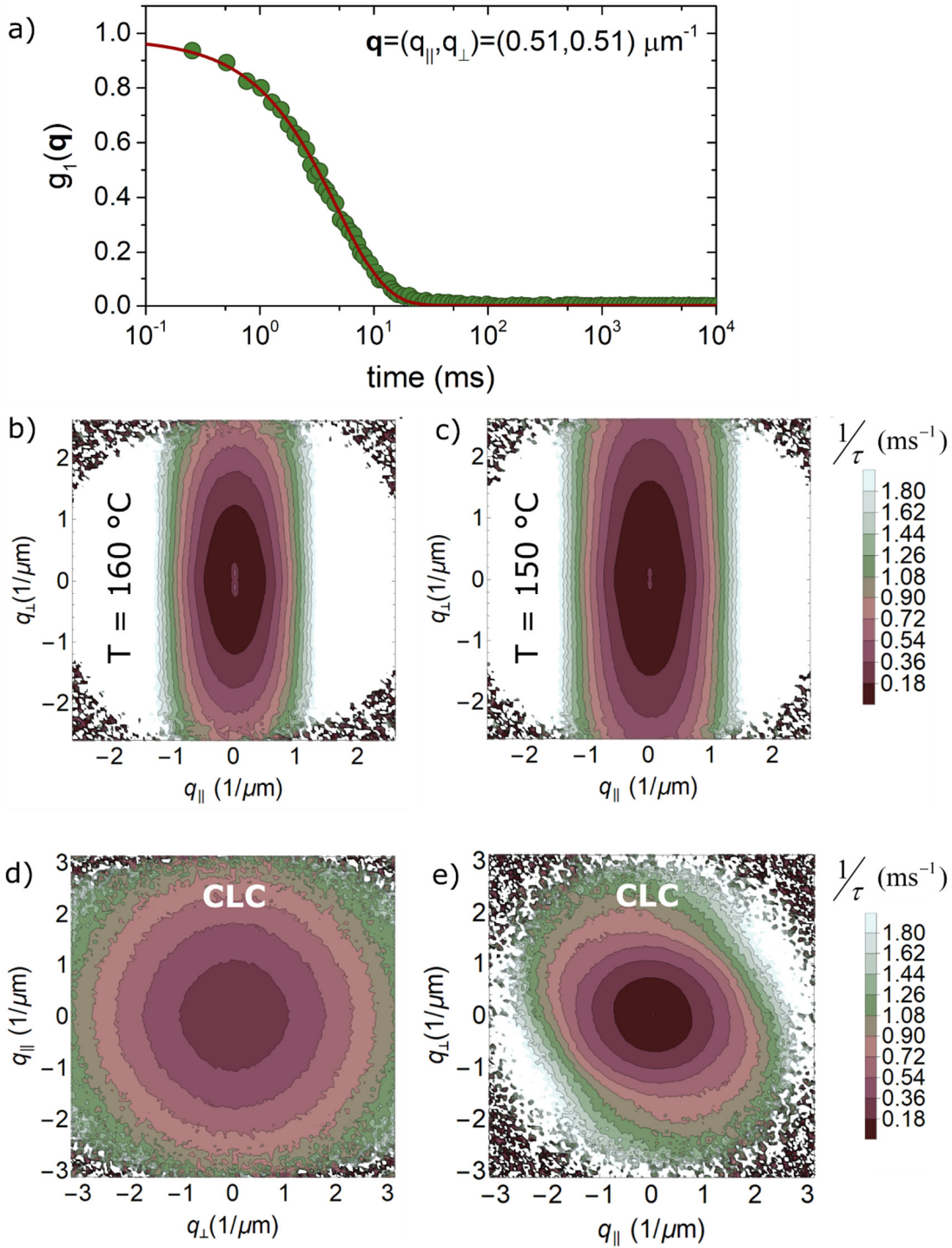

Fig. ESI.17. (Color online) cross-Differential dynamic microscopy examples in EE geometry. (a) Example of the image correlation function obtained in a c-DDM experiment for the measurement shown in the main manuscript Fig. 9 and its corresponding fit. (b-c) Relaxation rates for RM734 at two temperatures of the N phase. The comparison of both contour plots shows the slowing down of fluctuations while cooling. (d-e) Relaxation rates obtained for cholesteric liquid crystal (E7(Merck)+S-811(Merck))(d) with 4.7 μm pitch (S-811 1.92 wt%) in a 20 μm thick planar cell and (e) for a CLC with 20 μm pitch (S-811 0.41 wt%) in a wedge cell in the central area between the first and the second Grandjean-Cano lines.

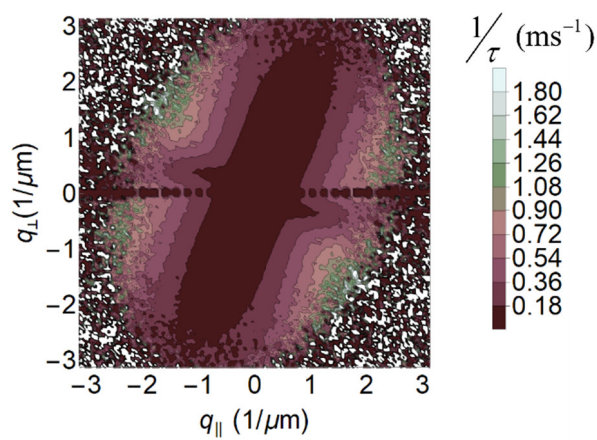

Fig. ESI.18 (Color online) cross-Differential dynamic microscopy example in EE geometry of the q-dependence of fluctuations for RM734 at 120 °C in an Instec IPS 9 µm cell with the alignment layer rubbed perpendicularly to the electrodes.

K) Schematic representation of possible emergence of biaxiality

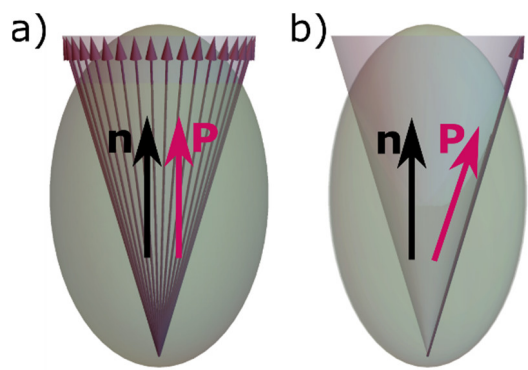

Fig. ESI.19. (Color online) Visualization of possibility for the emergence of biaxiality. (a) Dipole moments (maroon arrows) are randomly oriented on the cone. Ellipsoid represents the orientationaly averaged polarizability tensor, which is in this case uniaxial. (b) Dipole moments are preferentially oriented at certain angle on the cone e.g. determined by the field. Orientationaly averaged polarizability tensor is here biaxial. Averages of polarizability tensor are done assuming $P_1=1$.

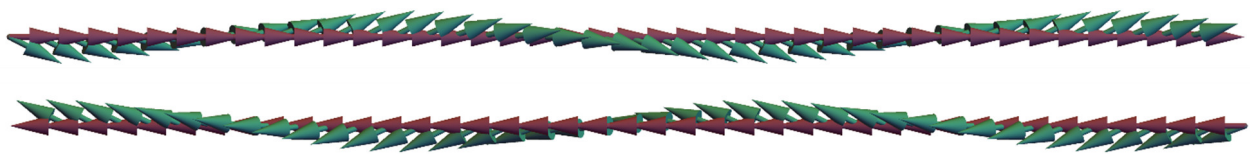

Fig. ESI.19 (Color online) Schematic representation of a polar heliconical structure. Maroon arrows represent the polarization P while green arrows represent the director.

## L) Index and description of Videos

- **Video 1**: Transition on cooling – EHC cell pi-twist domains.
- **Video 2**: Transition on cooling on Instec IPS cell with alignment rubbing parallel to electrodes.
- **Video 3**: Transition on cooling on an Instec IPS cell with alignment rubbing perpendicular to electrodes.
- **Video 4:** Deformation of pi-twist domains under the application of out of plane electric fields
- **Video 5:** 1 second reconstruction during electro-optic observations.

**Supporting Information Video 1 Description:**

$N$-$N_S$ phase transition recorded in an EHC 8.3 μm cell between crossed polarizers and cell rubbing direction slightly rotated with respect to the crossed polarizers. Image width corresponds to 775 μm. The video shows the destabilization of the homogeneous director orientation in the N phase, followed by stripped texture that gives rise again to a uniform texture, which finally relaxes to the final state through the propagation of a π-wall. At the end of the video, the imaged area is moved to visualize the propagation of the wall for the cell rubbing direction oriented along polarizers. Transition was recorded on cooling at 2 °C/min and has been speeded up by x4. Image width corresponds to 775 μm.

**Supporting Information Video 2 Description:**

$N$-$N_S$ phase transition recorded in an Instec IPS cell between crossed polarizers and cell rubbing direction (parallel to electrodes, SW-NE direction) rotated with respect to the crossed polarizers. The video shows and area where the formation of lens-shaped structures prevails after the transition stripes disappearance and how the structural relaxation propagates surrounding them, but not does not cross over such structures. Transition was recorded on cooling at 2 °C/min and has been speeded up by x2. Image width corresponds to 775 μm.

**Supporting Information Video 3 Description:**

$N$-$N_S$ phase transition recorded in an Instec IPS cell between crossed polarizers and cell rubbing direction (perpendicular to electrodes, NW-SE direction) rotated with respect to the crossed polarizers. Transition is characterized again by the distortion of the homogeneous nematic texture, followed by striped like texture which gives way to a homogeneous structure with lens-shaped structures in which structural relaxation observed in EHC and IPS cells with rubbing along electrodes does not take place. Transition was recorded on cooling at 2 °C/min and has been speeded up by x2. Image width corresponds to 1.2 mm.

**Supporting Information Video 4 Description:**

Application of out of plane fields in an EHC 5.4 μm cell at 120 °C. Initially the field is off. Then, a square-wave of 1 Hz and Vp= 0.5 V is applied, showing the pumping of the domain. Then an offset is applied and the voltage alternates between 0 V and 2Vp. At this field polarity the domain slowly grows. Then voltage is set to alternate between -2Vp and 0 V and the domain starts to shrink. Finally the field is turn off. Image width corresponds to 1.8 mm.

**Supporting Information Video 5 Description:**

Example of 1 Hz experiment reconstruction for (Top) short pulse and (Bottom) square-wave voltage profiles for Vp=2 V. Sample placed between crossed polarizers and with cell rubbing direction in vertical direction, i.e. perpendicular to the applied field. Gap between electrodes corresponds to 15 μm.